\documentclass[10pt]{revtex4-2}
\pdfoutput=1
\usepackage{graphicx}
\usepackage{epstopdf}
\usepackage{subfig}
\usepackage{amsmath}
\usepackage{verbatim}
\usepackage{mathrsfs}
\usepackage{amsfonts}
\usepackage{latexsym}
\usepackage{color}
\usepackage{units}
\usepackage{envmath}
\usepackage{natbib}
\usepackage{ctable}
\usepackage{soul}
\usepackage{mathtools}
\begin{document}
\title{Gravitational baryogenesis in $F(R)$ gravity’s rainbow}
\author{Parviz Goodarzi}
\affiliation{ Department of Physics, Faculty of Basic Sciences, Ayatollah Boroujerdi University, Boroujerd, Iran} 

\begin{abstract}
We investigate the gravitational baryogenesis scenario within the context  of $F(R)$ gravity’s rainbow which incorporates both modified gravity and energy-dependent spacetime.
This study explores a mechanism for generating baryon asymmetry based on  the interaction between the derivative of the Ricci scalar $R$ and the baryon current within the framework of $F(R)$ gravity's rainbow.
The rainbow functions, arising from quantum gravity effects, modify the gravitational interaction and the Friedmann equations, leading to a distinct evolution of the baryon asymmetry compared to standard $F(R)$
 gravity. 
We analyze the conditions under which a viable baryon asymmetry can be produced, taking into account the constraints from cosmological observations and the specific form of the $F(R)$ function.
By examining the cosmological equations in the context of $F(R)$ gravity's rainbow, we obtain power-law solutions for these equations.
We also identify the decoupling temperature and the ratio of baryonic number to entropy density in this model, depending on the model's parameters.
First, we discuss the acceptable intervals of the model's parameters which are defined by constraints on the background quantity.
We note that the decoupling temperature and the ratio of baryon-to-entropy in these models depend on the value of the rainbow function.
We compare the predictions of this model with the existing observational data.
Our results suggest that $F(R)$ gravity’s rainbow provides a novel mechanism for gravitational baryogenesis, potentially explaining the observed baryon asymmetry in the universe while incorporating quantum gravity corrections.
\end{abstract}

\maketitle

\section{Introduction}

There is substantial evidence indicating that matter is more prevalent than antimatter in the universe such as the cosmic microwave background radiation \citep{WMAP:2003ivt,WMAP:2006bqn}, the abundance of the primordial light elements emission from matter-antimatter annihilation \citep{Cohen:1997ac} and the big bang nucleosynthesis (BBN) \citep{Cyburt:2015mya}. Explaining the observed baryon asymmetry of the Universe (BAU) remains one of the most compelling open problems in cosmology and particle physics. Moreover, the Sakharov conditions provide a framework for baryogenesis, but the Standard Model of particle physics fails to provide sufficient CP violation or departure from thermal equilibrium to generate the observed BAU  \citep{Sakharov:1967dj}. Several intriguing and relevant mechanisms for generating baryon asymmetry that meet the Sakharov conditions have been proposed. The initial suggestion in this area involves the out-of-equilibrium decay of a massive particle, such as a superheavy GUT gauge boson or Higgs boson, referred to as GUT baryogenesis \citep{Weinberg:1979bt}. Another mechanism, known as the Affleck-Dine scenario, pertains to the decay of flat directions in supersymmetric models \citep{Affleck:1984fy}. Additionally, the possibility of generating baryon asymmetry at the electroweak scale has been explored; these interactions conserve the total baryon and lepton number, which is then converted into baryon asymmetry at the electroweak scale. This process is termed lepto-baryogenesis \citep{Fukugita:1986hr}. Furthermore, spontaneous baryogenesis has been proposed as a mechanism for generating baryon asymmetry while in thermal equilibrium, thereby removing the need for $C$ and $CP$ violation \citep{Cohen:1987vi,Cohen:1991iu}.

Within the context of Supergravity, Davoudiasl and his collaborators put forward a mechanism for producing baryon asymmetry through spontaneous baryogenesis. This approach involves an interaction between the derivative of the Ricci scalar curvature $R$ and the baryon current $J_\mu$, represented as $J^\mu\partial_\mu R$, which leads to a dynamic violation of CPT in an expanding Universe. This phenomenon is known as gravitational baryogenesis \citep{Davoudiasl:2004gf}.
While electroweak baryogenesis within the Standard Model is insufficient to account for the observed BAU, the model of gravitational baryogenesis provides an intriguing alternative. These models exploit the interplay between quantum field theory in curved spacetime and particle physics. Gravitational baryogenesis offers an alternative mechanism, linking the generation of the baryon asymmetry to the dynamics of the early universe and the interplay between gravity and particle physics. A typical scenario involves a non-minimal coupling between the baryon number current and a derivative of the Ricci scalar or other curvature invariants. During the early universe's rapid expansion, this coupling can result in a net increase in baryon number. In a universe dominated by matter, the interaction between gravitation and the baryon current can successfully generate the appropriate level of baryon asymmetry. Conversely, in a radiation-dominated universe ($\omega = 1/3$), it becomes essential to take into account the interactions among massless particles \citep{Davoudiasl:2004gf}. The ratio of baryon number density $n_B$ to entropy density $s$ is derived from accurate measurements of the cosmic microwave background radiation, resulting in the following value \citep{Planck:2018vyg}:
\begin{equation}\label{1}
Y_B\equiv{n_B\over s}=(0.864\pm 0.016)\times 10^{-10}.
\end{equation}

In recent years, various alternative scenarios for extending this coupling have garnered significant attention from numerous authors.
The influence of the time-dependent equation of state on gravitational baryogenesis has been investigated in \citep{Sadjadi:2007dx}.
The concept of gravitational baryogenesis within the framework of modified $f(R)$ gravity \citep{Nojiri:2010wj,Capozziello:2011et} has been explored in \citep{Lambiase:2006dq,Ramos:2017cot}. Modified gravity theories can successfully address the problem of baryon asymmetry generation in a radiation-dominated universe. This is due to the fact that the altered equations of motion can create notably different connections between scalar curvature and temperature. 
Also, gravitational baryogenesis in the context of $f(T)$ theory of gravity, which is a gravitational modification based on the torsion $T$ with the coupling $J^\mu\partial_\mu T$ and $J^\mu\partial_\mu f(-T)$ has been investigated \citep{Oikonomou:2016jjh}.
In \citep{Odintsov:2016hgc}, certain variant forms of gravitational baryogenesis are examined, specifically those that include the partial derivative of the Gauss-Bonnet scalar $\mathcal{G}$ coupled to the baryon current $J_\mu$ in the expression $J^\mu\partial_\mu \mathcal{G}$. Within the framework of Gauss-Bonnet baryogenesis, the ratio of baryon number density to entropy remains non-zero during the radiation-dominated era of the Universe.

Gravitational baryogenesis involving the term $J^\mu\partial_\mu  \mathcal{T}$ where $\mathcal{T}$ representing the trace of the energy-momentum tensor, in non-minimal $f(R,\mathcal{T})$ gravity has been investigated in \citep{Baffou:2018hpe}. In contrast to GR, they show that the baryon-to-entropy ratio is non-zero during the radiation-dominated era. Additionally, the authors of \citep{Sahoo:2019pat} explore the general form of the interaction $J^\mu \partial_\mu f(R, \mathcal{T}) $ within the framework of non-minimal $f(R, \mathcal{T})$ gravity.
 They found that the interaction proportional to $\partial_\mu \mathcal{T}$ led to unphysical outcomes.
The phenomenon of gravitational baryogenesis within the framework of $f(Q,\mathcal{T})$ gravity — where $Q$ represents nonmetricity — has been investigated, revealing results that align with observational data, as noted in \citep{Bhattacharjee:2020wbh}.

The concept of generalized gravitational baryogenesis is explored within the frameworks of $f(T,T_G)$ and $f(T,B)$ in the works of \citep{Azhar:2020coz,Nozari:2018ift}. Here, $T_G$ represents the teleparallel equivalent of the Gauss-Bonnet term, while $B = 2\nabla_\mu T^\mu$ denotes the boundary term that relates torsion to the Ricci scalar.
In these studies, the authors adopt a power-law form for the scale factor, represented as $a(t) = m_0 t^\gamma$, for each model. They derive the baryon-to-entropy ratio under the assumption that the universe is composed of a perfect fluid and dark energy.
The mechanism of gravitational baryogenesis, within the context of $f(Q,C)$ gravity, where $Q$ represents the non-metricity scalar and $C$ refers to the boundary term discussed in \citep{Usman:2024cya}. They demonstrate it through the coupling between the baryon matter current $J_\nu$ and the term $\partial_\nu(Q+C)$ generate observational results.
Moreover, the baryon asymmetry can be dynamically produced during an inflationary phase, driven by the creation of ultra-relativistic particles \citep{Lima:2016cbh}.
Gravitational baryogenesis in the non-minimal derivative coupling model, which is part of the Horndeski framework, has been analyzed in the high friction limit \citep{Goodarzi:2023ltp}. The findings indicate that gravitational baryogenesis can be explored at both low and high reheating temperatures.

In this paper, we will explore gravitational baryogenesis within the framework of $F(R)$ gravity's rainbow.
In recent years, the concept of gravity's rainbow has garnered significant interest from theorists seeking to explain various phenomena, including inflation, structure formation, warm inflation, dark energy, and the thermodynamics of black holes \citep{Ashour:2019wup,Shahjalal:2018hid,Upadhyay:2018vfu,Gim:2017rmn,Sefiedgar:2017tnt}.

Gravity's rainbow offers an intriguing perspective on a small-scale, ultraviolet modification of general relativity, interpreting general relativity as its low-energy limit \citep{Magueijo:2002xx,Awad:2013nxa}.
Like other theories of quantum gravity, its foundations are connected to the non-renormalizability of General Relativity and the difficulties associated with quantizing gravity \citep{Stelle:1976gc,Amelino-Camelia:2008aez,Gorji:2016laj}.
Gravity’s rainbow distinctively is a simple generalization of the double spatial relativity by taking into account the gravitational effects \citep{Rovelli:2008cj}.
Furthermore, in gravity's rainbow, the metric tensor is affected by the energy of the particle moving through space-time, especially in the vicinity of the Planck scale \citep{Magueijo:2002xx}.
Therefore, Gravity's rainbow incorporates a modified version of the general theory of relativity by taking length effects into account.
It is essential to emphasize that, in the low energy regime, the results of Gravity's rainbow should be consistent with standard general relativity \citep{Ling:2006az}.
The conventional dispersion relation in spatial relativity, given by $E^2-p^2=m^2$, is expected to be modified to deformations of dispersion relations, given by $E^2 \tilde{f}^2(E)-p^2 \tilde{g}^2(E)=m^2$. In this revised equation, $\tilde{f}(E)$ and $\tilde{g}(E)$ are known as rainbow functions, which arise in various theories of quantum gravity at energy levels close to the Planck length \citep{Magueijo:2002xx}. These functions must meet specific criteria in the low-energy infrared limit; in particular, as $E/M \to 0$, it is crucial that $\tilde{f}(E/M) \to 1$ and $\tilde{g}(E/M) \to 1$, where $M$ represents the energy scale at which quantum gravitational effects become significant.
The geometry of space-time in the context of gravity's rainbow is influenced by the energy of the test particles. As a result, each test particle with varying energy experiences a distinct geometry of space-time. This phenomenon gives rise to a set of metrics, known as a rainbow metric, which is defined by the energy $E$ and serves to describe the background of space-time rather than relying on a single metric. Therefore, in Gravity's Rainbow, the modified FLRW metric can be represented as:
\begin{equation}\label{1.1}
ds^2(E)=-{dt^2\over \tilde{f}^2(E)}+{a(t)^2\over \tilde{g}^2(E)}\delta_{ij}dx^idx^j,
\end{equation}
where $a(t)$ is a scale factor.
Additionally, research on $F(R)$ gravity within the context of an energy-dependent spacetime metric, known as $F(R)$ gravity's rainbow, which adds more degrees of freedom, allowing for more complex cosmological dynamics, has already been conducted in the literature.
Inflation and primordial fluctuation in $F(R)$ gravity's rainbow have been considered in the case that the rainbow function is written in the power-law form of the Hubble parameter \citep{Leyva:2022zhz,Waeming:2020rir}.
Starobinsky model and deformed Starobinsky model of inflation in context of rainbow gravity have been considered in \citep{Chatrabhuti:2015mws,Channuie:2019kus,Yuennan:2024lkz}.
 $F(R)$ gravity's rainbow bridges general relativity and quantum gravity by incorporating low-energy quantum components \citep{Burton-Villalobos:2025tgc}. Meanwhile, gravitational baryogenesis, a topic within particle physics, encourages us to explore the concept of gravitational baryogenesis within the framework of $F(R)$ gravity's rainbow.
 
The paper is organized as follows:
In Section 2, we provide a brief introduction to $F(R)$ gravity's rainbow. We derive the field equations for this framework and obtain power-law solutions for two scenarios: when the rainbow functions are equal ($\tilde{f} = \tilde{g}$) and when the rainbow function $\tilde{g}$ is equal to one ($\tilde{g}=1$).
In Section 3, we explore gravitational baryogenesis within the framework of $F(R)$ gravity's rainbow. We derive the decoupling temperature $T_D$ and the baryon-to-entropy ratio $Y_B$ as functions of the model's parameters.   
In Section 4, we perform a numerical analysis of gravitational baryogenesis in the framework of $F(R)$ gravity's rainbow. We examine the quantitative impacts of baryon asymmetry and compare our findings with observational data.
In the last section, we conclude our results.

We use natural units $\hbar=c=1$ throughout the paper.

\section{$F(R)$ gravity's rainbow}

In this section, we will examine power-law solutions in $F(R)$ gravity within a rainbow background metric. We will focus on two scenarios: first, when $\tilde{f} = \tilde{g}$, and second, when the rainbow function $\tilde{g}$ is equal to one ($\tilde{g} = 1$). Modifications to general relativity are anticipated to occur in the very early universe, where some corrections to general relativity may emerge at high curvature. One of the simplest forms of modified gravity theory is known as $F(R)$ gravity \citep{Sotiriou:2008rp,DeFelice:2010aj}. We begin by deriving the field equations for $F(R)$ gravity using the rainbow metric, as described in \citep{Waeming:2020rir}. To accomplish this, we will examine the full action of $F(R)$ gravity, taking into account the effects of gravitational baryogenesis as 
\begin{equation}\label{2}
S={1\over 2\kappa^2}\int F(R)\sqrt{-g}d^4x+S_{M}+S_{B},
\end{equation}
where we have defined $\kappa^2=8\pi G$, $F(R)$ is an arbitrary  function of Ricci scalar $R$, $g$ is determinant of the metric,  $S_{M}$ is the matter action.
To explain gravitational baryogenesis, we define the action $S_B$ as the interaction between the derivative of the Ricci scalar curvature $\partial_\mu R$ and the baryon current $J^\mu$ \citep{Davoudiasl:2004gf}.
 \begin{equation}\label{3}
S_{B}={1\over M_{*}^2}\int{d^4x\sqrt{-g}(\partial_\mu R)J^\mu},
\end{equation}
where $M_{*}$ is the cutoff scale of the effective theory.
The energy-momentum tensor of the matter can be derived by varying the matter action with respect to the metric, as shown below
\begin{equation}\label{4}
T^{(M)}_{\mu\nu}={-2\over\sqrt{-g}}{\delta(\sqrt{-g}\mathcal{L}_M)\over \delta g^{\mu\nu}},
\end{equation}
where the energy-momentum tensor of the matter field satisfies the continuity equation such that $\nabla^\mu T^{(M)}_{\mu\nu}=0$.
In the conventional approach, it is important to highlight that the energy-momentum tensor of matter is modeled as a perfect fluid in the following form:
\begin{equation}\label{5}
T^{(M)}_{\mu\nu}=(\rho+P)u_{\mu}u_{\nu}+Pg_{\mu\nu},
\end{equation}
where $u^{\mu}$ represents the four-velocity of the matter field, while $ \rho$ and $P$ denote the energy density and pressure of the universe, respectively.
The field equation can be derived by varying the action \eqref{2} with respect to metric $g_{\mu\nu}$
\begin{equation}\label{6}
F_{,R}(R)R_{\mu\nu}-{1\over2}F(R)g_{\mu\nu}-\nabla_{\mu}\nabla_{\nu}F_{,R}(R)+g_{\mu\nu}\square F_{,R}(R)=\kappa^2 T_{\mu\nu}^{(M)},
\end{equation}
where $F_{,R}(R)=\partial F(R)/\partial R$ and the operator $\square$ is defined as $\square\equiv (1/\sqrt{-g})\partial_\mu(\sqrt{-g}g^{\mu\nu}\partial_\nu)$. We will now examine the field equation \eqref{6} within the context of the rainbow metric, which takes the following form:
\begin{equation}\label{7}
ds^2=-{1\over \tilde{f}^2(t)}dt^2+{1\over \tilde{g}^2(t)}a^2(t)\delta_{ij}dx^idx^j.
\end{equation}
By applying the rainbow metric \eqref{7}, we can express the Ricci scalar in the following manner 
\begin{equation}\label{8}
R=6\tilde{f}^2\Big(\dot{H}+2H^2-4H{\dot{\tilde{g}}\over\tilde{g}}+3{\dot{\tilde{g}}^2\over\tilde{g}^2}-{\ddot{\tilde{g}}\over\tilde{g}}\Big)+6\tilde{f}\dot{\tilde{f}}\Big(H-{\dot{\tilde{g}}\over\tilde{g}}\Big),
\end{equation}
where $H=\dot{a}/a$ is the Hubble parameter, overdot sign is derivative with respect to cosmic time $t$.
The energy density can be written as
\begin{equation}\label{9}
\kappa^2\rho=\tilde{f}^2\Big[3\big(H^2 F_{,R}+H\dot{F}_{,R}\big)+\dot{F}_{,R} {\dot{\tilde{f}}\over\tilde{f}}-6F_{,R}H{\dot{\tilde{g}}\over\tilde{g}}+3F_{,R} {\dot{\tilde{g}}^2\over\tilde{g}^2}-3\dot{F}_{,R}{\dot{\tilde{g}}\over\tilde{g}}\Big]-{1\over2}\big(F_{,R} R-F\big).
\end{equation}
Additionally, the pressure is expressed as
\begin{eqnarray}\label{10}
&&3F_{,R}H^2-3\dot{F}_{,R} H+3F_{,R} \dot{H}+3F_{,R} H{\dot{\tilde{f}}\over\tilde{f}}-\dot{F}_{,R}{\dot{\tilde{f}}\over\tilde{f}}
-4F_{,R}{\dot{\tilde{g}}^2\over\tilde{g}^4}  \nonumber \\
&&+6F_{,R} H{\dot{\tilde{g}}\over\tilde{g}^3}-3\dot{F}_{,R} {\dot{\tilde{g}}\over\tilde{g}^3}+F_{,R}{\dot{\tilde{g}}\over\tilde{g}^3}+F_{,R} {\dot{\tilde{f}}\over\tilde{f}}{\dot{\tilde{g}}\over\tilde{g}^3}-3F_{,R} H^2{1\over \tilde{g}^2} \nonumber \\
&&+2\dot{F}_{,R} H {1\over\tilde{g}^2}+\ddot{F}_{,R}{1\over\tilde{g}^2}-F_{,R} \dot{H}{1\over\tilde{g}^2}+6F_{,R}{\dot{\tilde{g}}^2\over\tilde{g}^2}
-F_{,R}H{\dot{\tilde{f}}\over\tilde{f}}{1\over\tilde{g}^2} \nonumber \\
&&+\dot{F}_{,R}{\dot{\tilde{f}}\over\tilde{f}}{1\over\tilde{g}^2}-6F_{,R} H {\dot{\tilde{g}}\over\tilde{g}}+3\dot{F}_{,R}{\dot{\tilde{g}}\over\tilde{g}}-3F_{,R} {\ddot{\tilde{g}}\over\tilde{g}}-3F_{,R} {\dot{\tilde{f}}\over\tilde{f}}{\dot{\tilde{g}}\over\tilde{g}}\nonumber\\
&&-F{(g-1)(g+1)\over2\tilde{f}^2\tilde{g}^2}=
-{\kappa^2\over\tilde{f}^2\tilde{g}^2}\big(\rho\tilde{g}^2+P\big). 
\end{eqnarray}
We choose to explicitly parameterize the rainbow function using the power-law representation of the Hubble parameter, specifically \citep{Chatrabhuti:2015mws}:
\begin{equation}\label{11}
\tilde{f}^2=1+\Big({H\over M}\Big)^{2\alpha},
\end{equation}
with the rainbow parameter $\alpha > 0$. Furthermore, in the context of late-time cosmology, we have $\tilde{f} \approx 1$. During the inflationary period, we focus on the situation where $H^2 \gg M^2$. As a result, in our study, the rainbow function can be expressed as $\tilde{f} \approx \left(H/M\right)^\alpha$.
In the following, we explore the power-law expansion of the universe, characterized by the scale factor $a(t)=At^\gamma$, where the power-law index $\gamma$ is greater than one ($\gamma > 1$).
In this context, the Hubble parameter $H$ can be represented as $H = {\gamma}/{t}$, and the rainbow function $\tilde{f}$ transforms into
\begin{equation}\label{12}
\tilde{f}=\Big({\gamma\over M}\Big)^\alpha t^{-\alpha}.
\end{equation}
Additionally, we can express $\dot{\tilde{f}}/\tilde{f}=-\alpha t^{-1}$ as a function of cosmic time $t$.
In the majority of the literature analyzing rainbow gravity, two scenarios are usually taken for simplicity: one in which the rainbow function is defined as $\tilde{f} = \tilde{g}$ and the other where $\tilde{g} = 1$. In the following subsections, we will derive the solutions for the power-law form of the scale factor in two different scenarios.

\subsection{Case 1: $\tilde{f}=\tilde{g}$}

\begin{figure}[t]
\begin{center}
       \scalebox{0.55}{\includegraphics{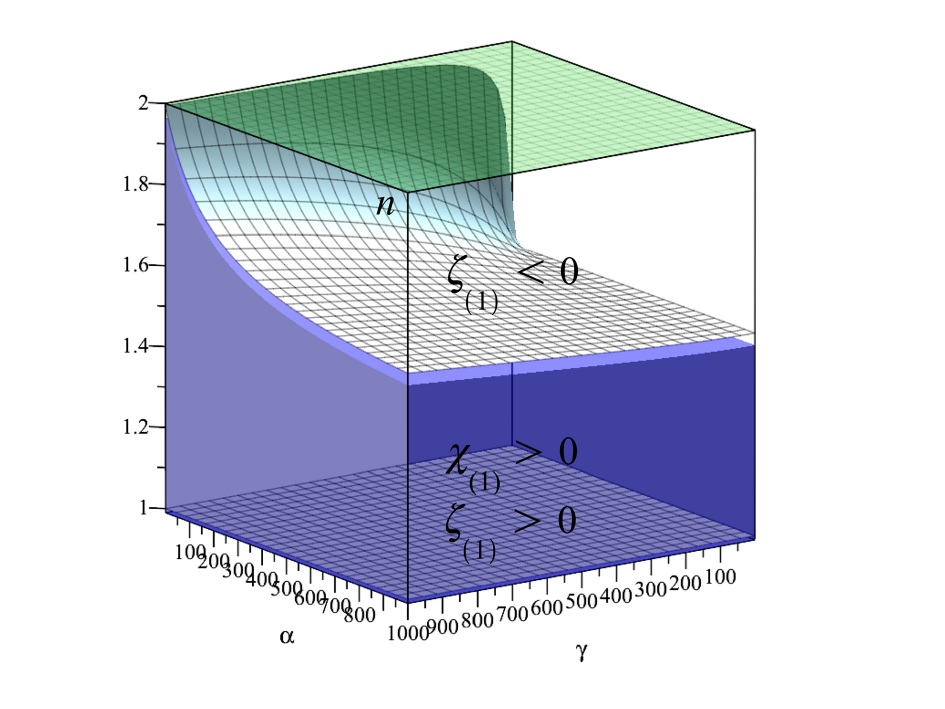}}
    \scalebox{0.55}{\includegraphics{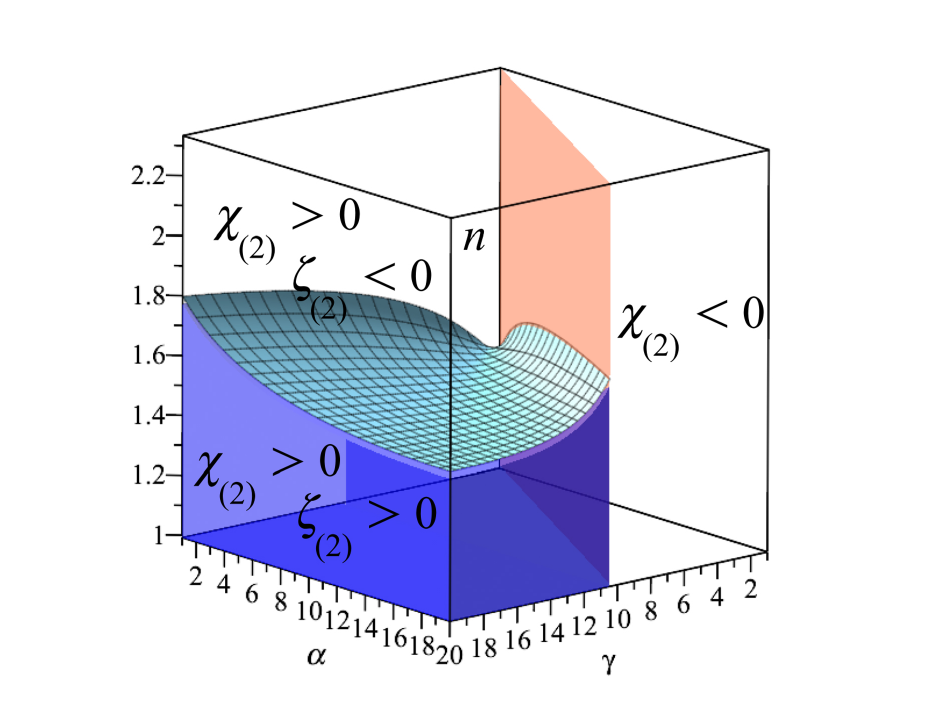}}
   \end{center}
    \caption{\footnotesize The surface described by the equation $\mathcal{S}(n,\alpha,\gamma)=0$ is represented in the coordinate system $(\alpha,\gamma,n)$. On the left, we illustrate case 1, where $\tilde{f}=\tilde{g}$, and on the right, we show case 2, where $\tilde{g}=1$. The permissible values for $n$, $\alpha$, and $\gamma$ are highlighted in the area where $\{\chi,\zeta\}> 0$.}
  \label{fig1}
\end{figure}

A convenient choice for rainbow functions is $\tilde{f} = \tilde{g}$. With this simplification, and utilizing equations \eqref{8} and \eqref{12}, the Ricci scalar $R$ can then be calculated as a function of cosmic time in the following manner
\begin{equation}\label{13}
R=\chi_{(1)} t^{-2(\alpha+1)},
\end{equation}
where $\chi_{(1)}$ in this expression takes the form of
 \begin{equation}\label{14}
\chi_{(1)}=6\Big({\gamma\over M}\Big)^{2\alpha}\big(\gamma+\alpha\big)\big(\alpha+2\gamma-1\big).
\end{equation}
In the cases where $\alpha=1-2\gamma$, we observed that the Ricci scalar either vanishes or takes on constant values for $\alpha=-1$ showing no asymmetry between baryons and antibaryons.
In this study, we consider the functional $F(R)$ in a straightforward form
\begin{equation}\label{15}
 F(R)=BR^n,
\end{equation}
where $B$ and $n$ are the constant parameters of the model.
As a result $F_{,R}=nBR^{n-1}$. Also, the first and second time derivatives of this function are then expressed as follows:
\begin{eqnarray}\label{16}
\dot{F}_{,R}&=&(n-1){\dot{R}\over R}F_{,R}, \\
\ddot{F}_{,R}&=&(n-1)\Bigg({d\over dt}\Big({\dot{R}\over R}\Big)+(n-1) \Big({\dot{R}\over R}\Big)^2\Bigg) F_{,R}.
\end{eqnarray}
By applying relation \eqref{13} for the Ricci scalar, we can express these relations in the following form:
\begin{eqnarray}\label{17}
F_{,R}&=&nB\chi^{(n-1)} t^{-2(n-1)(\alpha+1)}, \\
\label{17.1}
\dot{F}_{,R}&=&-2(\alpha+1)(n-1) t^{-1} F_{,R}, \\
\label{17.2}
\ddot{F}_{,R}&=&-2(\alpha+1)(n-1)\Big(1+2(n-1)(\alpha+1)\Big)t^{-2}F_{,R}.
\end{eqnarray} 
By performing some calculations and applying the relations (\ref{13}, \ref{17}, \ref{17.1}, \ref{17.2}) within the equation \eqref{9}, we can derive the energy density of the Universe as a function of cosmic time in the following manner 
\begin{equation}\label{18}
\rho=\zeta_{(1)} t^{-2n(\alpha+1)},
\end{equation}
where the quantity $\zeta_{(1)}$ in this equation can be expressed as
\begin{eqnarray}\label{19}
\zeta_{(1)}=\kappa^2\Big({\gamma\over M}\Big)^{2\alpha} B\chi_{(1)}^{(n-1)}\Big[3n(\alpha+\gamma)^2&-&2n(n-1)(\alpha+1)(3\gamma+2\alpha) \nonumber \\
&-&3(n-1)(\gamma+\alpha)(\alpha+2\gamma-1)\Big].
\end{eqnarray}
If we substitute $\chi_{(1)}$ from relation \eqref{14} into the equation \eqref{19} we have
\begin{eqnarray}\label{20}
\zeta_{(1)}&=&\kappa^2\Big({\gamma\over M}\Big)^{2n\alpha} B \Big[6(\gamma+\alpha)(\alpha+2\gamma-1)\Big]^{(n-1)}
\Big[3n(\alpha+\gamma)^2 \nonumber \\
&-&2n(n-1)(\alpha+1)(3\gamma+2\alpha)
-3(n-1)(\gamma+\alpha)(\alpha+2\gamma-1)\Big].
\end{eqnarray}
As mentioned in the previous sections, we have $\{\alpha,\gamma\}>0$. Furthermore, for the expansion of the universe, it is necessary that
 $\{R,\rho\}>0$, which leads to the conclusion that $\{\chi_{(1)},\zeta_{(1)}\}>0$. The condition $\chi_{(1)}>0$ implies that $\gamma>{1-\alpha\over2}$.
To determine where the quantity $\zeta_{(1)}>0$, we define $\mathcal{S}_{(1)}(\alpha,\gamma,n)$ as follows:
\begin{eqnarray}\label{20.1}
\mathcal{S}_{(1)}(\alpha,\gamma,n)=3n(\alpha+\gamma)^2 
-2n(n-1)(\alpha+1)(3\gamma+2\alpha)
-3(n-1)(\gamma+\alpha)(\alpha+2\gamma-1).
\end{eqnarray}
Next, we need to identify the domain where $\mathcal{S}_{(1)}$ is positive. Thus, the equation $\mathcal{S}_{(1)}(\alpha,\gamma,n)=0$ defines a surface in the $(\alpha,\gamma,n)$ coordinate system that serves as a boundary, distinguishing areas where $\mathcal{S}_{(1)}$ is positive from those where it is negative.
As a result, the surface defined by $\mathcal{S}_{(1)}=0$ becomes
\begin{eqnarray}\label{20.2}
&n&=\Big(-3\gamma^2+3\alpha\gamma+4\alpha^2+9\gamma+7\alpha \nonumber \\
&&+\sqrt{64\alpha^4+8(7+30\gamma)\alpha^3+(297\gamma^2+210\gamma+1)\alpha^2+6(21\gamma^3+42\gamma^2+\gamma)\alpha+9\gamma^4+90\gamma^3+9\gamma^2} \Big) \nonumber \\
&& /4(\alpha+1)(2\alpha+3\gamma).
\end{eqnarray} 
Additionally, it has been noted that as either $\gamma$ or $\alpha$ in relation \eqref{20.2} tends toward infinity, the value of $n$ tends toward 2 or $3/2$ respectively 
\begin{eqnarray}\label{20.3}
Lim_{\alpha\rightarrow\infty} n={3\over2},~~~ \space Lim_{\gamma\rightarrow\infty} n=2.
\end{eqnarray}
Thus, if we select $n \geq 2$, the sign of the quantity $\mathcal{S}_{(1)}$ does not change for any value of $\gamma$ and $\alpha$. The acceptable ranges for the quantities $\alpha$, $\gamma$, and $n$ are depicted in Figure \ref{fig1}.

\subsection{Case 2: $\tilde{g}=1$}

Another convenient choice for rainbow functions is $\tilde{g}=1$ \citep{Leyva:2022zhz,Waeming:2020rir}. We can now compute the Ricci scalar $R$ as a function of cosmic time using equations \eqref{8} and \eqref{12} follow as
\begin{equation}\label{21}
R=\chi_{(2)} t^{-2(\alpha+1)},
\end{equation}
where $\chi_{(2)}$ in this relation is in the following form
 \begin{equation}\label{22}
\chi_{(2)}=6\Big({\gamma\over M}\Big)^{2\alpha}\gamma \big(2\gamma-\alpha-1\big).
\end{equation}
For the simple $F(R)$ described in equation \eqref{15}, the energy density in this scenario is expressed as
\begin{equation}\label{23}
\rho=\zeta_{(2)} t^{-2n(\alpha+1)},
\end{equation}
where the quantity $\zeta_{(2)}$ can be written as
\begin{eqnarray}\label{24}
\zeta_{(2)}=\Big({\gamma\over M}\Big)^{2\alpha} B\chi_{(2)}^{(n-1)}\Big[3n\gamma^2&-&2n(n-1)(\alpha+1)(3\gamma-\alpha) \nonumber \\
&-&3\gamma(n-1)(2\gamma-\alpha-1)\Big].
\end{eqnarray}
By substituting $\chi_{(2)}$ from relation \eqref{22} into the equation \eqref{24} we have
\begin{eqnarray}\label{25}
\zeta_{(2)}&=&\Big({\gamma\over M}\Big)^{2n\alpha} B \Big[6\gamma(2\gamma-\alpha-1)\Big]^{(n-1)}
\Big[3n\gamma^2 \nonumber \\
&-&2n(n-1)(\alpha+1)(3\gamma-\alpha)
-3\gamma(n-1)(2\gamma-\alpha-1)\Big].
\end{eqnarray}
Additionally, in the spatial case where $\tilde{g}=1$, the equation \eqref{9} simplifies to the following form:
\begin{eqnarray}\label{26}
\ddot{F}_{,R}-H\dot{F}_{,R}+2F_{,R}\dot{H}+2F_{,R} H{\dot{\tilde{f}}\over\tilde{f}}=-{\kappa^2\over\tilde{f}^2}\big(\rho+P\big). 
\end{eqnarray}
Thus, by substituting equations (\ref{13}, \ref{17}, \ref{17.1}, \ref{17.2}) into equation \eqref{26}, the pressure can be written as
\begin{equation}\label{27}
P=\xi_{(2)} t^{-2n(\alpha+1)},
\end{equation}
where $\xi_{(2)}$ is expressed in the following manner
\begin{eqnarray}\label{28}
\xi_{(2)}&=&-\Big({\gamma\over M}\Big)^{2n\alpha}B\chi_{(2)}^{(n-1)}\kappa^{-2} \Big[2n(\alpha+1)(n-1)\big[-3\gamma+(\alpha+1)(2n-1)\big] \nonumber \\ &+& 2n\gamma(\alpha+1)(n-2)+3n\gamma^2-3\gamma(n-1)(2\gamma-\alpha-1)\Big].
\end{eqnarray}
Consequently, the equation of state is formulated as
\begin{eqnarray}\label{29}
\omega={P\over \rho}={\xi_{(2)}\over\zeta_{(2)}}&=&-\Big[2n(\alpha+1)(n-1)\big[-3\gamma+(\alpha+1)(2n-1)\big] 
+ 2n\gamma(\alpha+1)(n-2)\nonumber \\ 
&+&3n\gamma^2-3\gamma(n-1)(2\gamma-\alpha-1)\Big]\Bigg/\Big[3n\gamma^2-2n(n-1)(\alpha+1)(3\gamma-\alpha) \nonumber \\
&-&3\gamma(n-1)(2\gamma-\alpha-1)\Big].
\end{eqnarray}

Such as case 1 in previous subsection we need $\{\chi_{(2)},\zeta_{(2)}\}>0$. The condition $\chi_{(2)}>0$ implies that $\gamma>{1+\alpha\over2}$.
To identify the conditions under which the quantity $\zeta_{(2)}>0$, we define $\mathcal{S}_{(2)}(\alpha,\gamma,n)$ in the following manner:
\begin{eqnarray}\label{29.1}
\mathcal{S}_{(2)}(\alpha,\gamma,n)=3n\gamma^2-2n(n-1)(\alpha+1)(3\gamma-\alpha)
&-&3\gamma(n-1)(2\gamma-\alpha-1).
\end{eqnarray}
To determine the region where $\mathcal{S}_{(2)}$ is positive, we need to find the roots of the equation $\mathcal{S}_{(2)}(\alpha,\gamma,n)=0$. This equation defines a surface in the $(\alpha,\gamma,n)$ coordinate system, which acts as a boundary that separates the areas where $\mathcal{S}_{(2)}$ is positive from those where it is negative.
Consequently, the surface characterized by $\mathcal{S}_{(1)}=0$ is
\begin{eqnarray}\label{29.2}
&n&=\Big(-2\alpha^2+(9\gamma-2)\alpha-3\gamma^2+9\gamma \nonumber \\
&&+\sqrt{9\gamma^4+90(\alpha+1)\gamma^3+9(-3\alpha^2-2\alpha+1)\gamma^2 -12\alpha(\alpha+1)^2\gamma+4\alpha^2(\alpha+1)^2} \Big) \nonumber \\
&& /4(\alpha+1)(3\gamma-\alpha).
\end{eqnarray} 
Also, we have been noted that as either $\gamma$ or $\alpha$ tends toward infinity, the value of $n$ tends toward 2 or $0$ respectively
\begin{eqnarray}\label{29.3}
Lim_{\alpha\rightarrow\infty} n=0,
~~~ \space Lim_{\gamma\rightarrow\infty} n=2.
\end{eqnarray}
Therefore, when we choose $n \geq 2$, the sign of the quantity $\mathcal{S}_{(2)}$ remains constant regardless of any values of $\gamma $ and $\alpha$. The permissible ranges for $\alpha$, $\gamma$, and $n$ are illustrated in Figure \ref{fig1}.

\begin{figure}[t]
\begin{center}
  \scalebox{1}{\includegraphics{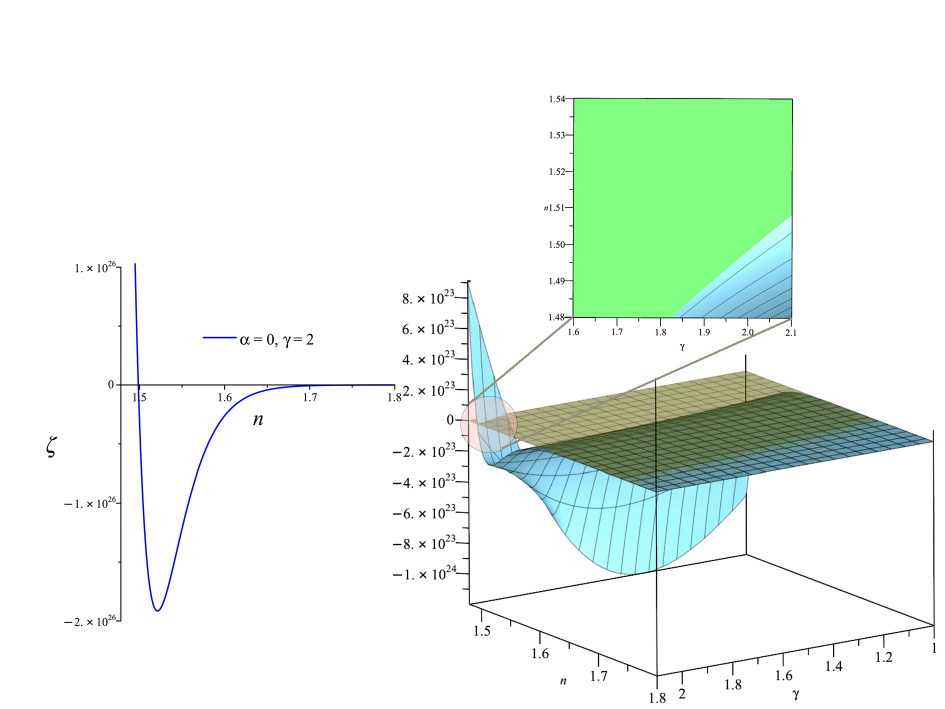}}
   \end{center}
    \caption{ \footnotesize The 3-dimensional plot on the right illustrates $\zeta$ as a function of $\gamma$ and $n$, with the parameter $\alpha$ held constant at 0.1. On the left, the solid blue curve represents the relationship between $\zeta$ and $n$ for fixed values of $\alpha = 0.1$ and $\gamma = 2$. The region where $\zeta > 0$ indicates the feasible values for $n$ and $\gamma$.}
\label{fig2}
\end{figure}

\begin{figure}[t]
\begin{center}
  \scalebox{1}{\includegraphics{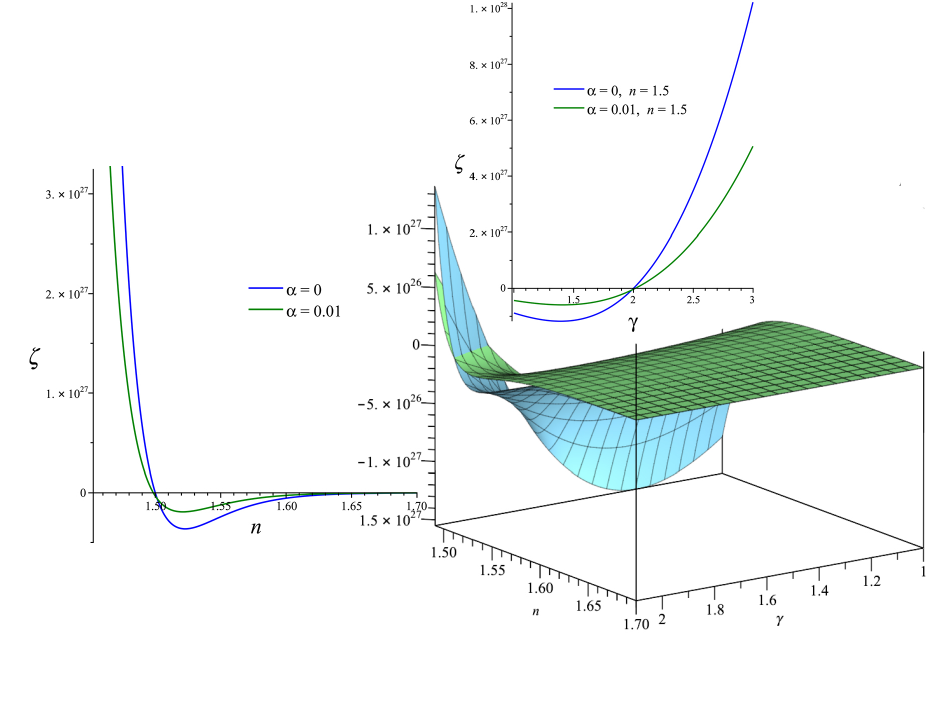}}
   \end{center}
    \caption{ \footnotesize The 3D plot on the right illustrates $\zeta$ as a function of $\gamma$ and $n$ for two constant parameter values: $\alpha=0$ and $\alpha=0.01$. The solid blue and green curves on the left-hand side represent the intersection of this 3D plot with the surface of constant $\gamma=2$, corresponding to $\alpha=0$ and $\alpha=0.01$, respectively. Similarly, the solid blue and green curves in the upper right panel depict the intersection of the 3D plot with the surface of constant $n=1.5$, also for $\alpha=0$ and $\alpha=0.01$.}
  \label{fig3}
\end{figure}

\begin{figure}[t]
\begin{center}
  \scalebox{1}{\includegraphics{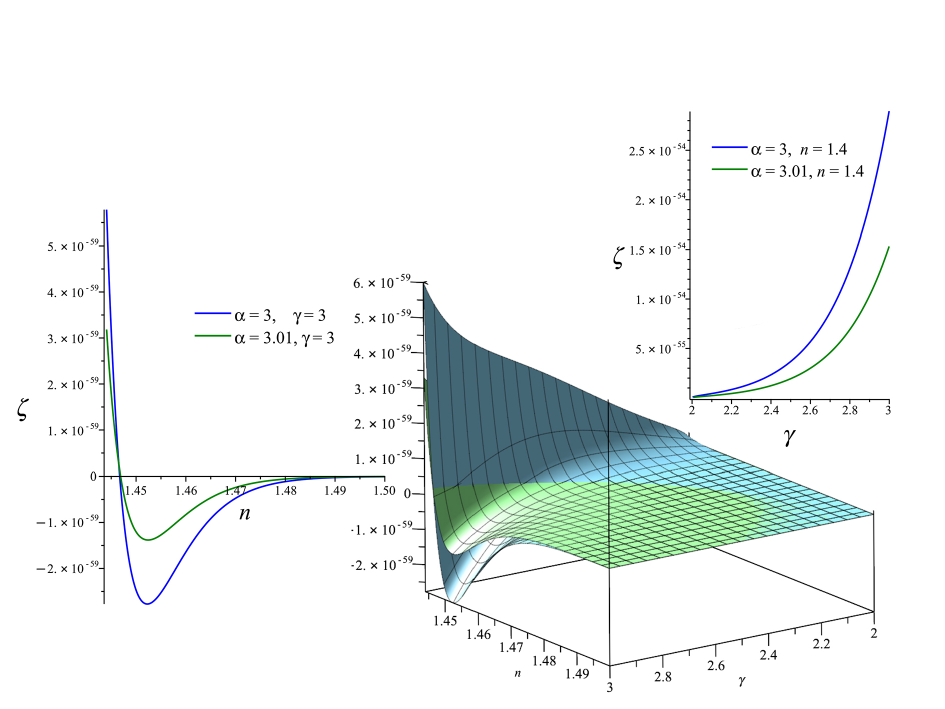}}
    \end{center}
    \caption{ \footnotesize The 3D-plot on the right illustrates $\zeta$ as a function of $\gamma$ and $n$ for two constant parameter values: $\alpha=3.0$ and $\alpha=3.01$. The solid blue and green curves on the left-hand side represent the intersection of this 3D plot with the surface of constant $\gamma=3$, corresponding to $\alpha=3.00$ and $\alpha=3.01$, respectively. Similarly, the solid blue and green curves in the upper right panel depict the intersection of the 3D plot with the surface of constant $n=1.4$, also for $\alpha=0$ and $\alpha=0.01$.}
  \label{fig4}
\end{figure}

\section{Gravitational baryogenesis}

In the previous section, we observed that the Ricci scalar $R$ and the energy density of the Universe $\rho$ can be expressed as power-law functions of cosmic time in two different scenarios. Therefore, we can represent $R$ and $\rho$ in the following manner:
 \begin{eqnarray}\label{29.1}
R&=&\chi t^{-2(\alpha+1)},\\
\label{29.2}
\rho &=&\zeta t^{-2n(\alpha+1)}.
\end{eqnarray}
In this context, $\chi$ may refer to either $\chi_{(1)}$ or $\chi_{(2)}$, while $\zeta$ can denote either $\zeta_{(1)}$ or $\zeta_{(2)}$.
This study explores the impact of rainbow gravity on baryon asymmetry by evaluating the ratio of baryon number to entropy. In a state of thermal equilibrium, B-violating interactions can create a net baryon asymmetry. Within the context of an expanding Universe, we refer to action (\ref{1}) as described in \citep{Davoudiasl:2004gf}.
\begin{equation}\label{30}
 {1\over M_\star}(\partial_\mu R)J^\mu={\dot{R}\over M_{\star}^2}(g_bn_b+g_{\bar{b}}n_{\bar{b}}),
\end{equation}
where $g_b=-g_{\bar{b}}$ denotes the number of intrinsic degrees of freedom of baryons. $n_b$ and $n_{\bar{b}}$ are the number densities of baryon and antibaryon respectively. An effective chemical potential follow as $\mu_b=-\mu_{\bar{b}}=g_b\dot{R}/M_{\star}^2$, the entropy density of the Universe is given by $s=2\pi^2g_\star {\tau}^{3}/45$, the baryon number density, in thermal equilibrium, becomes $n_B=(g_bn_b+g_{\bar{b}}n_{\bar{b}})=-g_b\mu_b\tau^2/6$ \citep{Kolb:1990vq}, where $\tau$ is the temperature of the Universe.
As a result, we can write the baryon-to-entropy ratio (baryon asymmetry) in an accelerating universe as
\begin{equation}\label{31}
Y_B\equiv {n_B\over s}\approx -{15g_b^2\over4\pi^2 g_\star}{\dot{R}\over M_{\star}^2\tau}\Big|_{\tau=\tau_{D}},
\end{equation}
where the temperature $\tau_D$ is the temperature of the Universe at which the baryon current violation decouples.
Now, with the time derivative of Ricci scalar curvature (\ref{29.1}) we have
\begin{equation}\label{32}
\dot{R}=-2\chi(\alpha+1)t^{-2\alpha-3},
\end{equation}
To calculate the decoupling temperature as a function of decoupling time we assume that the energy density of the Universe is described with $\rho=\pi^2g_{\star}\tau^4/30$. Now, by using relation \eqref{23} we have
\begin{equation}\label{33}
t_D=\Big({\pi^2\over 30 g_{\star}\zeta}\Big)^{-1/2n(\alpha+1)} \tau_D^{-2/ n(\alpha+1)}.
\end{equation} 
Finally, by substituting $\dot{R}$ and $\rho$ from equations (\ref{32}) and (\ref{29.2}) into equation (\ref{31}), we obtain baryon asymmetry $Y_B$ as a function of decoupling temperature as
\begin{equation}\label{34}
Y_B\equiv {n_B\over s}=-{15g_b\chi(\alpha+1)\over 2\pi^2 g_{\star} M_{\star}^2} \Big({\pi^2\over30 g_{\star}\zeta}\Big)^{(2\alpha+3)/2n(\alpha+1)}\tau_D^{2(2\alpha+3)/n(\alpha+1)-1}.
\end{equation}

We continue with a brief mention of the origin of the B-violating interaction that is indispensable for any baryogenesis scenario. We assume that B-violating interactions, which are given by an operator $\mathcal{O}_B$ of mass dimension $D=4+n$ \citep{Davoudiasl:2004gf,Fukushima:2016wyz}. We need $n>0$ for the B-violating interaction. In the B-violating interactions, coupling constants are proportional to $M_B^{-n}$, where $M_B$ is the mass scale, the rate of generation of B-violating interaction in thermal equilibrium with the temperature $\tau$ can be cast in the form \citep{Davoudiasl:2004gf}
\begin{equation}\label{35}
\Gamma_B={\tau^{2m+1}\over M_B^{2m}}.
\end{equation}
Decoupling of B-violating processes occurs at $\tau=\tau_D$, when $\Gamma$ falls below $H={\gamma/t}$. Therefore, we can obtain the temperature of decoupling from equation \eqref{29.2} as
\begin{equation}\label{37}
t_D=\gamma M_B^{2n} \tau_D^{-(2m+1)}.
\end{equation}
Consequently, we can obtain decoupling temperature in the following form 
\begin{equation}\label{38}
\tau_D=\Big(\gamma M_B^{2n}\Big)^{n(\alpha+1)\over n(\alpha+1)(2m+1)-2} \Big({\pi^2\over30\zeta g_{\star}}\Big)^{1\over 2n(\alpha+1)(2m+1)-4)}.
\end{equation}
Finally, by substituting the decoupling temperature $\tau_D$ from relation \eqref{38} into equation \eqref{34}, we obtain
\begin{equation}\label{39}
Y_B=-{15 g_b\chi(\alpha+1)\over 2\pi^2 g_{\star} M_{\star}^2} \Big(\gamma M_B^{2n}\Big)^{2(2\alpha+3)-n(\alpha+1)\over n(\alpha+1)(2m+1)-2} \Big({\pi^2\over 30\zeta} g_{\star}\Big)^{(2m+1)(2\alpha+3)-1)\over 2[n(\alpha+1)(2m+1)-2]}.
\end{equation}

From the condition $H^2/M^2\gg1$ we can constrain the models parameters
\begin{equation}\label{40}
\tau_D^{(2m+1)}\gg M M_B^{2m}.
\end{equation}

\begin{figure}[t]
\begin{center}
  \scalebox{0.33}{\includegraphics{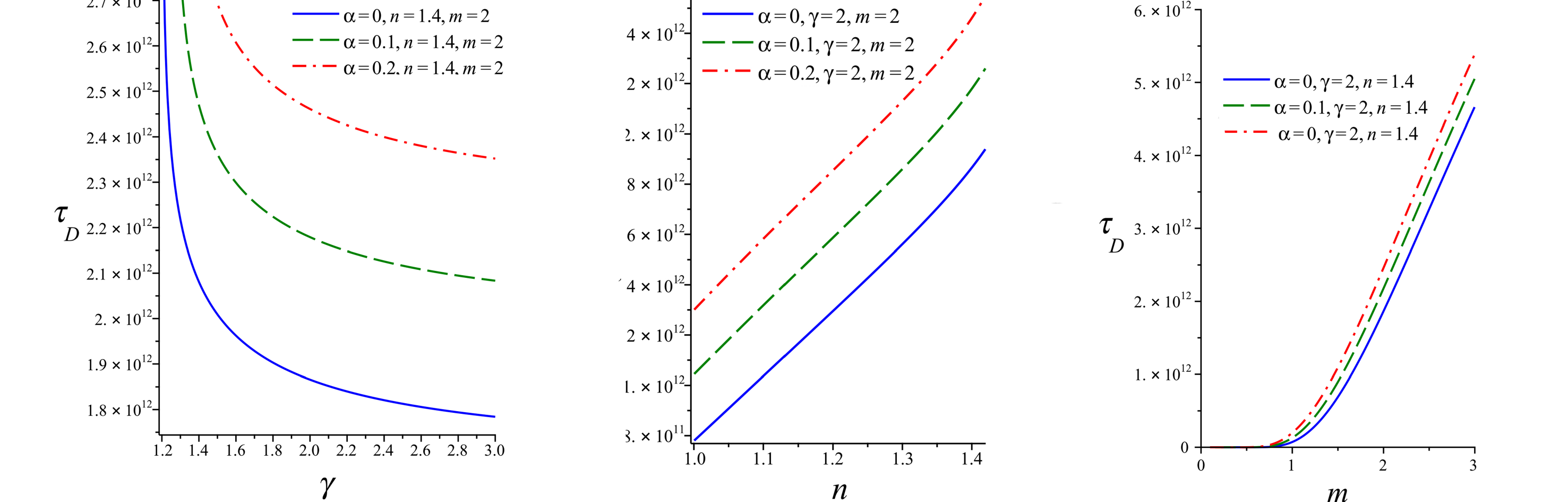}}
  \scalebox{0.32}{\includegraphics{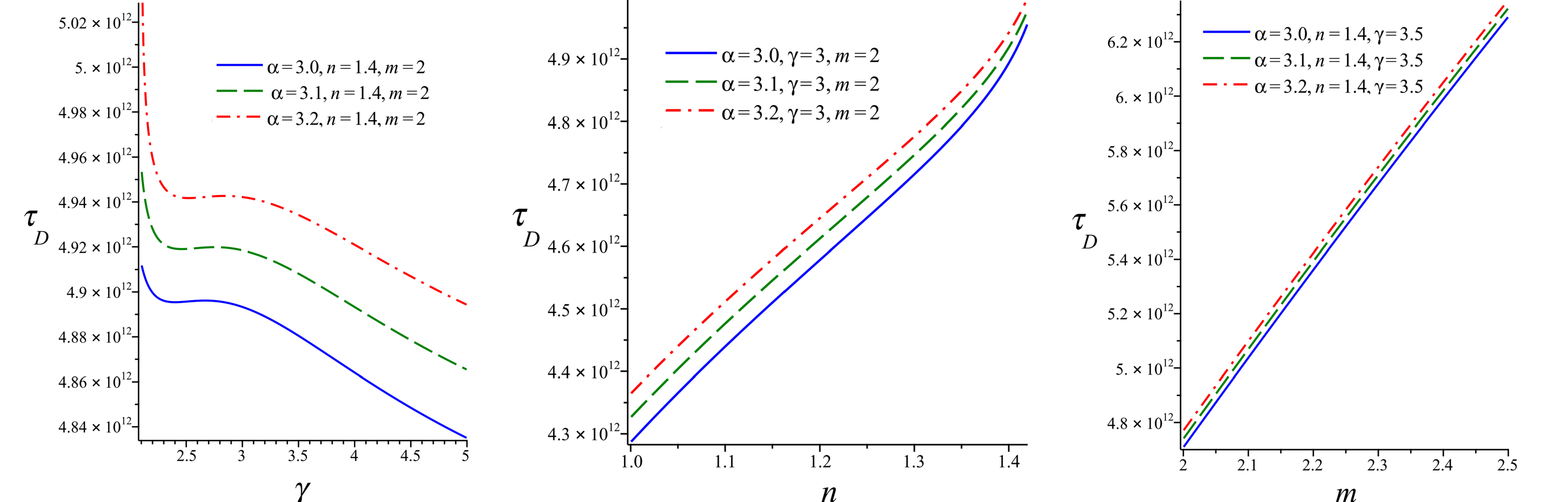}}
    \end{center}
    \caption{ \footnotesize We depict the behavior of the decoupling temperature $\tau_D$ in terms of $\gamma$, $n$ and $m$ for several different values of the parameter $\alpha$. In these plots we assume that $M=10^{-8}M_p$ and $M_B=10^{-5}M_p$.}
  \label{fig5}
\end{figure}

\begin{figure}[t]
\begin{center}
  \scalebox{0.32}{\includegraphics{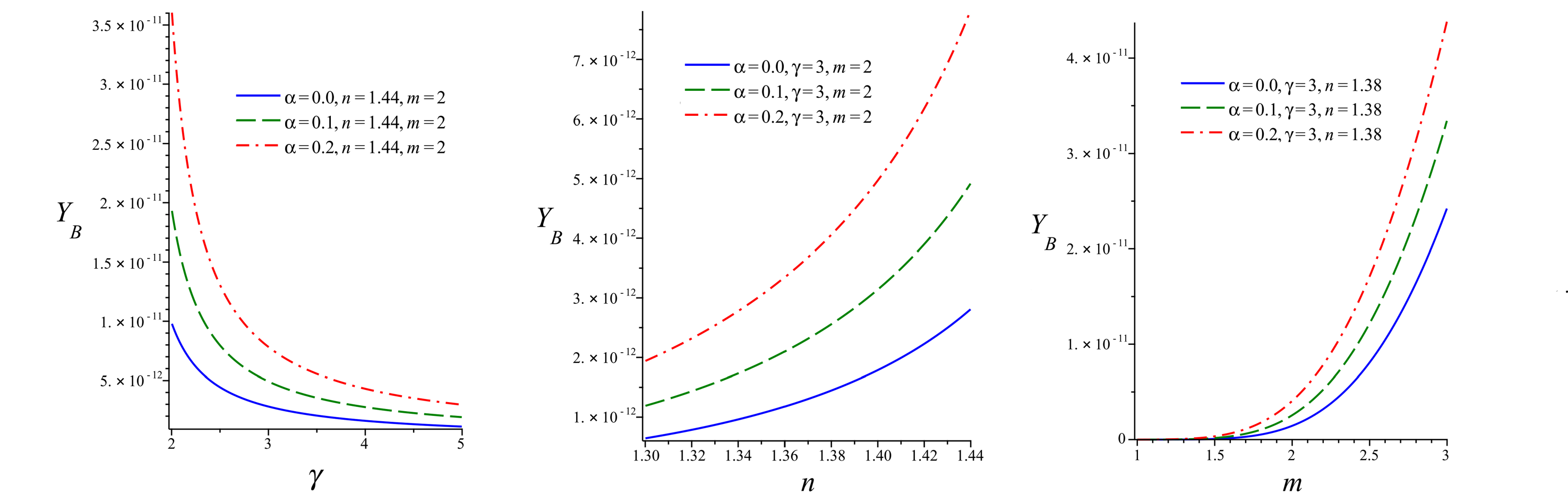}}
  \scalebox{0.32}{\includegraphics{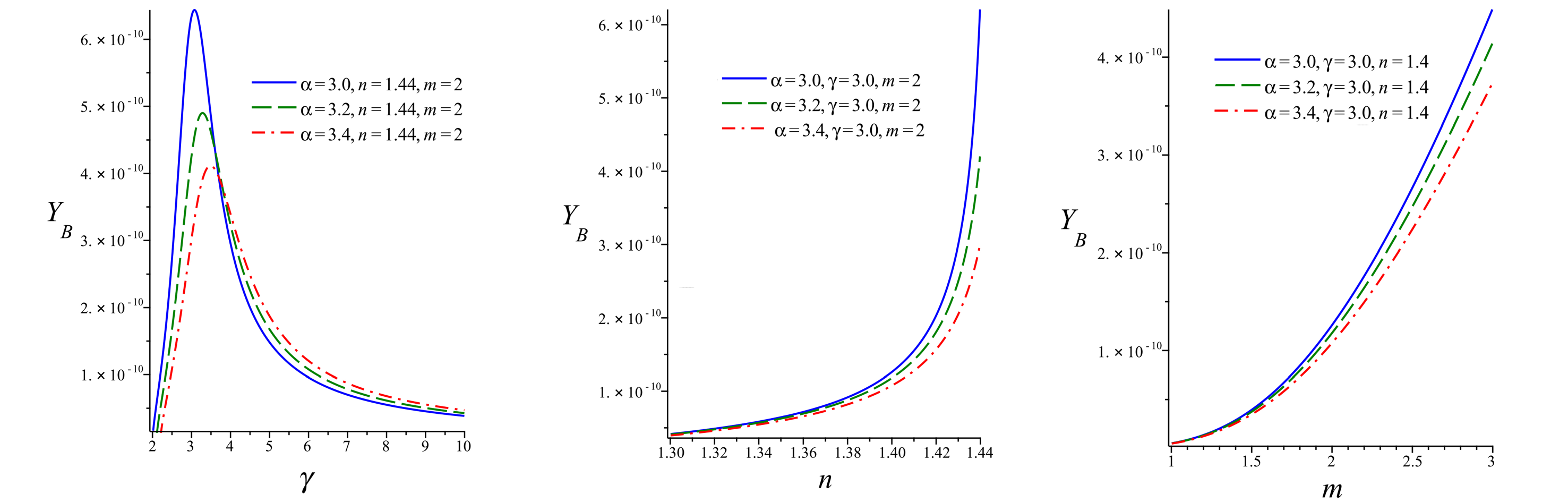}}
       \end{center}
    \caption{ \footnotesize We depict the behavior of the ratio of baryon-to-entropy $Y_B$ in terms of $\gamma$, $n$ and $m$ for several different values of the parameter $\alpha$. In these plots we assume that $M=10^{-8}M_p$ and $M_B=10^{-5}M_p$.}
  \label{fig6}
\end{figure}

\begin{figure}[t]
\begin{center}
  \scalebox{0.37}{\includegraphics{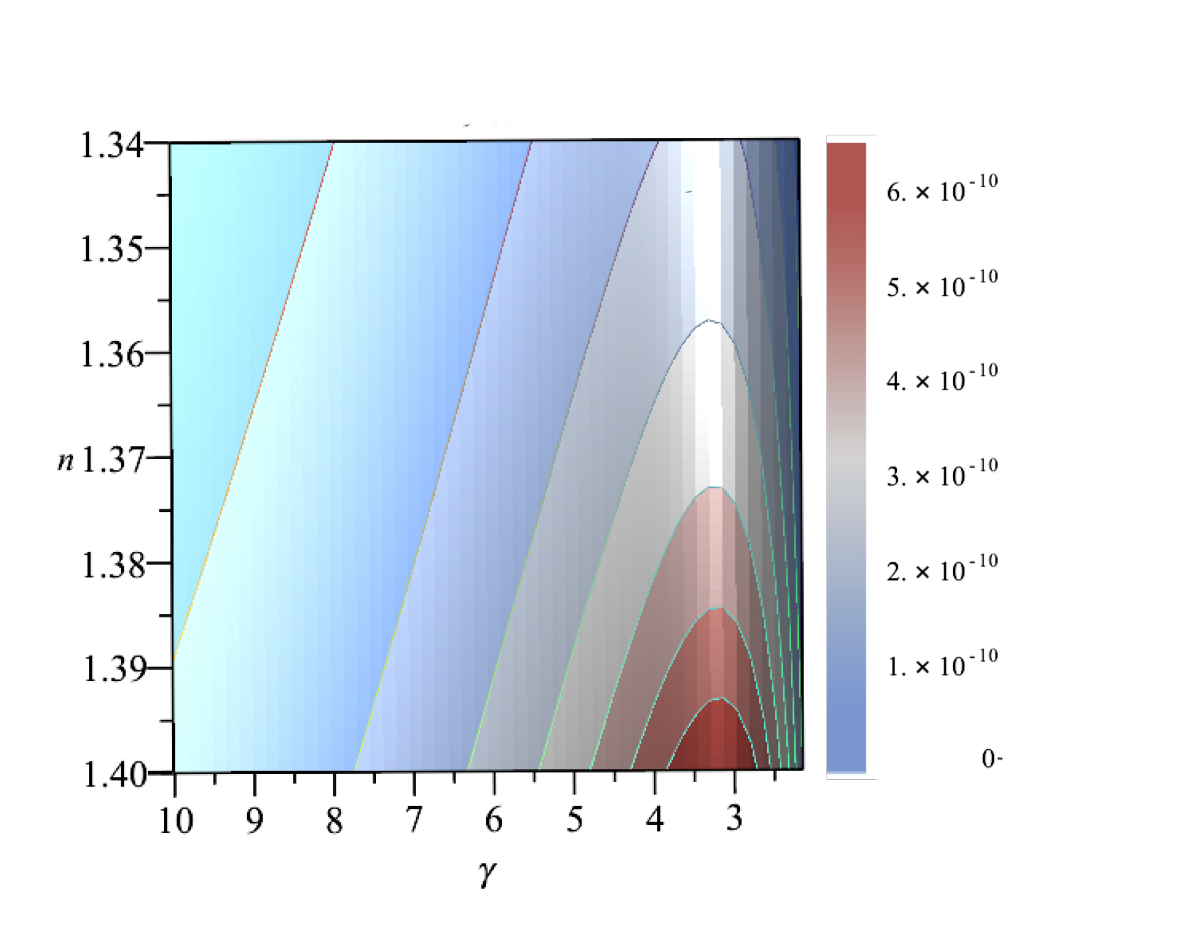}}
  \scalebox{0.37}{\includegraphics{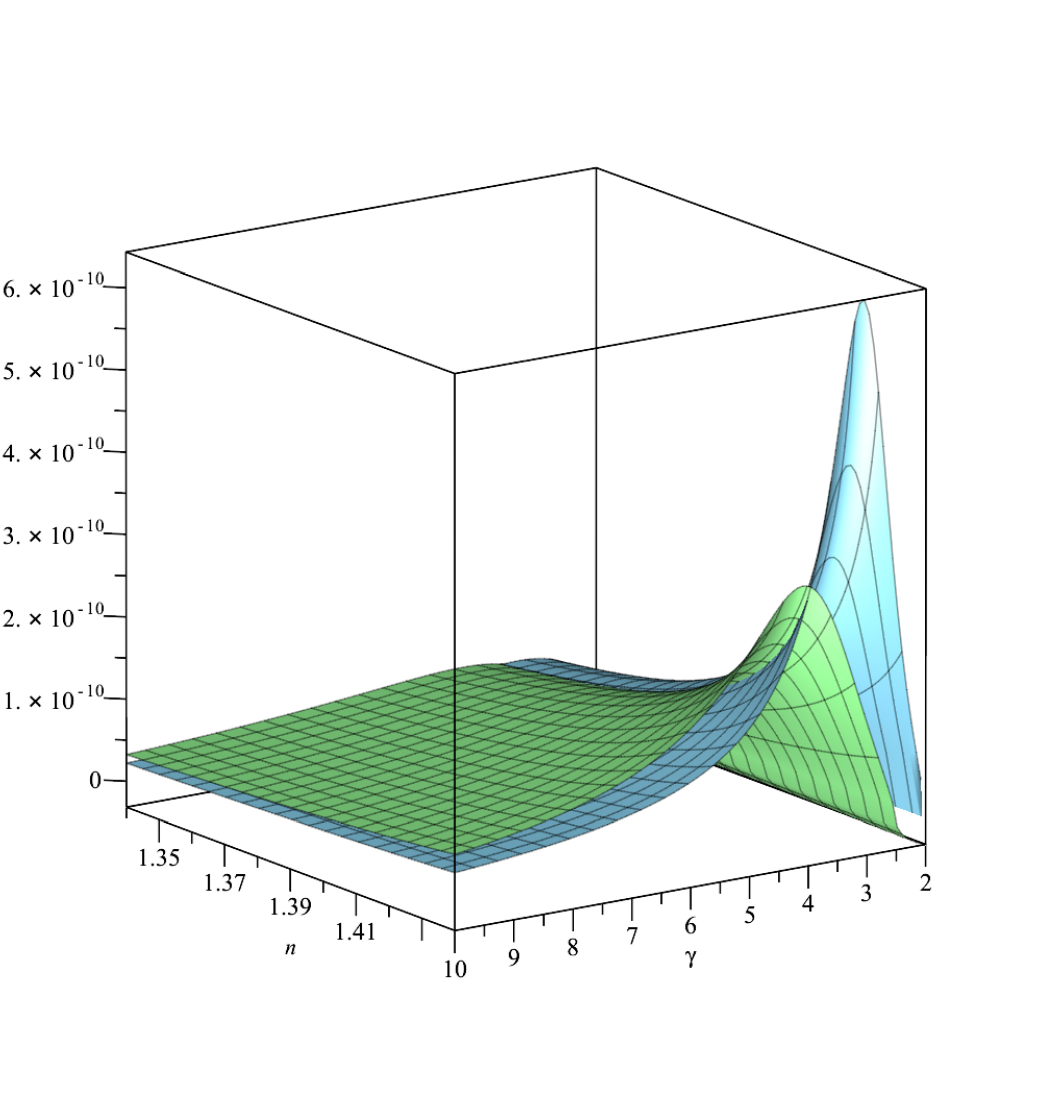}}
       \end{center}
    \caption{ \footnotesize We depict the behavior of the ratio of baryon-to-entropy $Y_B$ in terms of $\gamma$, $n$ with $m=2$ for two constant  values of the parameter $\alpha=3$ and $\alpha=3.01$. In these plots we assume that $M=10^{-8}M_p$ and $M_B=10^{-5}M_p$.}
  \label{fig7}
\end{figure}
 
\begin{figure}[t]
\begin{center}
  \scalebox{0.5}{\includegraphics{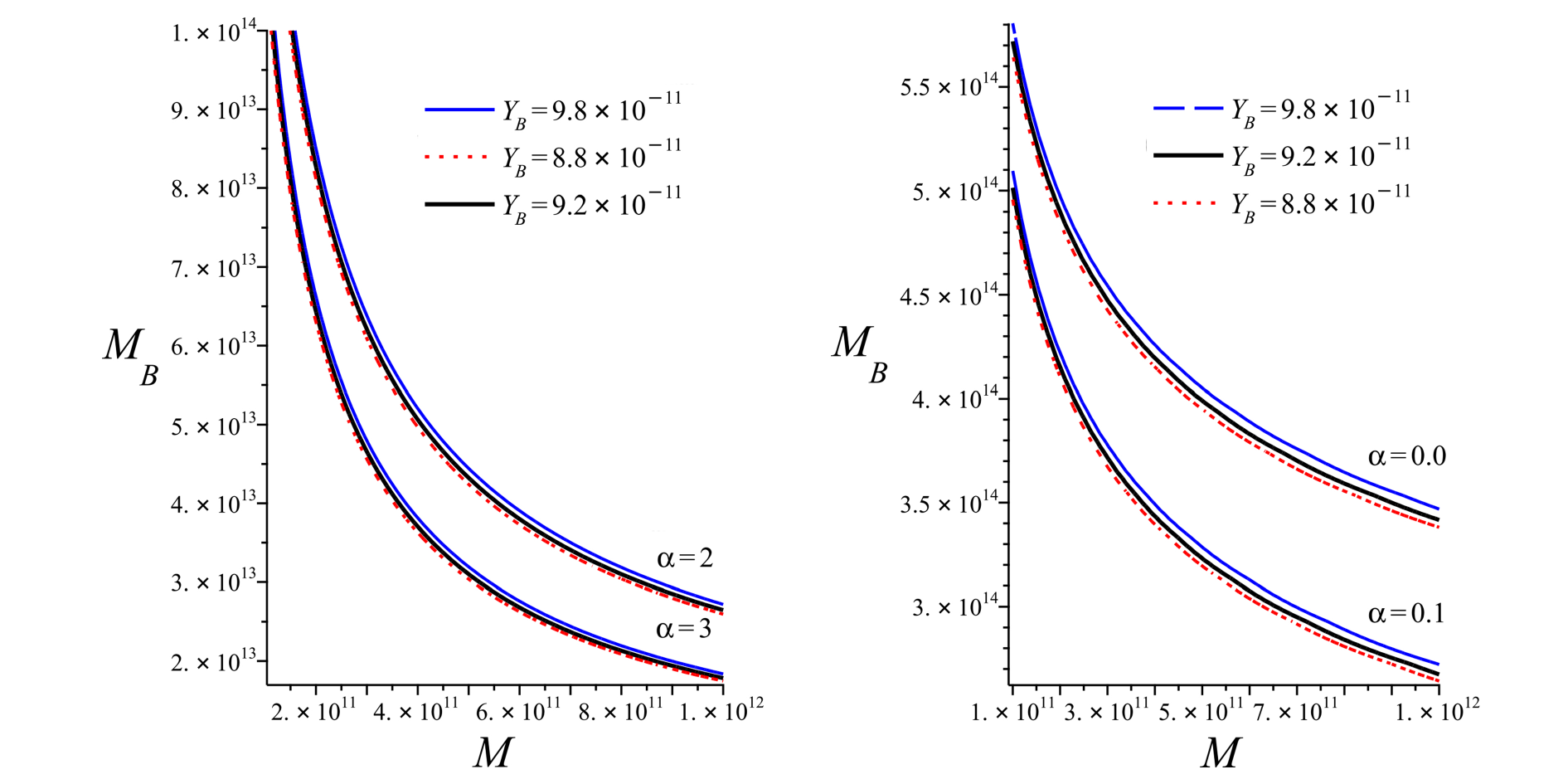}}
       \end{center}
    \caption{ \footnotesize The acceptable values for $M$ and $M_B$ that account for the observed baryon asymmetry $Y_B = 9.2 \times 10^{-11}$ are illustrated. The solid black curve represents a dimension-6 B-violating interaction $(m = 2)$. The red dashed curve and the blue dotted curve correspond to the error rates in the ratio of baryon density to entropy.}
  \label{fig8}
\end{figure}
 
\section{numerical analysis}

In the previous section, we explored exact solutions for inflation and gravitational baryogenesis within the framework of $F(R)$ gravity's rainbow.
Our analysis yielded the Ricci scalar, energy density of matter, decoupling time, decoupling temperature, and baryon asymmetry as functions of the model parameters for both cases: $\tilde{f}=\tilde{g}$ and $\tilde{g}=1$, exactly. 
We used a power-law scale factor given by $a(t)=At^\gamma$, assumed that $\tilde{f}\approx (H/M)^\alpha$ during the inflationary period, and utilized a power-law functional form of $F(R)$ expressed as $F(R)=BR^n$ and the rate of generating B-violating interactions $\Gamma_B=\tau^{2m+1}/M_B^{2m}$.

We have identified nine parameters in the model, denoted as $A, \gamma, M, \alpha, B, n, M_B, m, M_{\star}$. 
In models with a large number of parameters, a key challenge is determining the appropriate values of these parameters.
Our goal is to determine these parameters in such a way that the ratio of baryon number to entropy is consistent with cosmological observations.

In the first step, we have investigated gravitational baryogenesis during inflation. As a result, based on the relationship $\ddot{a}/a = \gamma(\gamma-1)t^{-2}$, it is essential for $\gamma$ to be greater than $1$.
 
Additionally, the energy density of matter $\rho$ must be a positive value. According to relation \eqref{18}, the signature of the energy density of matter is influenced by three parameters $\gamma ,\alpha$ and $n$ within the parameter $\zeta$ in the form of $\rho=\zeta t^{-2n(\alpha+1)}$, therefore we need $\zeta>0$. Due to the nearly equivalent results of these two cases discussed in the previous sections,  we will focus on case 2, where $\tilde{g}=1$. For all subsequent investigations, we will focus solely on Case 1 $(\tilde{g} = 1)$, using $\zeta$ and $\chi$ to represent $\zeta_1$ and $\chi_1$ respectively.

In FIG. \ref{fig2}, we present $\zeta$ as a function of $\gamma$ and $n$ for fixed value of $\alpha=0$.
To emphasize the regions where $\zeta>0$, we present a plane section of the three-dimensional plot $\zeta$ at a fixed value of $\gamma=2$ to plot the level curve of $\zeta$ solely against $n$, on the left panel. Furthermore, the possible values for $n$ and $\gamma$ are represented by the blue area in the plot above.
In the scenario where $\alpha=0$ and $\gamma=2$, we observe that for $n<1.5$, the value of $\zeta$ is greater than 0.

To examine how the quantity of $\alpha$ influences the behavior of $\zeta$, we have illustrated this scenario for two values of $\alpha$: $\alpha=0$ and $\alpha=0.01$, as shown in FIG. \ref{fig3}.
To highlight the areas where $\zeta>0$, we provide a plane section of the three-dimensional plot of $\zeta$ at a fixed value of $\gamma=2$ to plot the level curve of $\zeta$ solely against $n$, for two different values of $\alpha$: $\alpha=0$ and $\alpha=0.01$, in the left panel. Additionally, we provide a plane section of the three-dimensional plot of $\zeta$ at a fixed value of $n=1.5$ to plot the level curve of $\zeta$ solely against $\gamma$, for two different values of $\alpha$: $\alpha=0$ and $\alpha=0.01$, in the above panel.
To demonstrate the impact of larger $\alpha$ values, we created Figure \ref{fig4} for $\alpha=3$ and $\alpha=3.01$, as depicted in Figure \ref{fig3}.
We have observed that when $n < 1.42$, with $\alpha = 3$ and $\gamma > 2$, both quantities $\zeta$ and $\chi$ are positive.
 
To examine how the decoupling temperature $\tau_D$ changes with respect to $\gamma$, $n$, and $m$, we provide the plots in Figure \ref{fig5}. In these plots, we assume $M=10^{-8}M_p$ and $M_B=10^{-5}M_p$. The upper panels of Figure \ref{fig5} display the plots for low values of the parameter $\alpha$, specifically $0$, $1.1$, and $0.2$, while the lower panels show high values of $\alpha$, which include $3.0$, $3.1$, and $3.2$.
For the parameters within the ranges $1< \gamma<3$, $1<n<1.4$, and $0<m<3$, as well as $\alpha \in \{0, 0.1, 0.2, 3, 3.1, 3.2\}$, the decoupling temperature $\tau_D$ is of order $10^{12} \, \text{GeV}$.
As you can see in Figure \ref{fig5}, The higher $\gamma$ values decrease the decoupling temperature, while higher $n$ and $m$ values increase it.
Therefore, if we want to explain, the generation of baryon asymmetry in high decoupling temperature, then we have to choose higher values of $n$ and $m$.

The behavior of the ratio of baryon-to-entropy $Y_B$ against $\gamma$, $n$, and $m$, are depicted in Figure \ref{fig6} for several values of $\alpha$, specifically $\{0, 0.1, 0.2, 3, 3.2, 3.4\}$.
As illustrated in Figure \ref{fig6}, higher values of $\gamma$ lead to a decrease in the baryon-to-entropy ratio $Y_B$, whereas increased values of $n$ and $m$ result in an increase in this ratio. 
The graph in Figure \ref{fig6} illustrates that the ratio of baryon-to-entropy, in relation to $\gamma$, reaches its peak around $\gamma=3$ for the parameters $\alpha=3$, $n=1.44$, and $m=2$. Additionally, as the values of $\alpha$ increase, the maximum value of this ratio decreases.
For all of these plots, we choose the appropriate values of model parameters such that $\{\chi, \zeta\}>0$.

In Figure \ref{fig7}, we present the baryon-to-entropy ratio as a function of $\gamma$ and $n$ in both three-dimensional plots (right panel) and contour plots (left panel) for two values of the parameter $\alpha$: 3 and 3.01. These plots reveal a maximum values of plots around $\gamma=3$.

Figure \ref{fig8} illustrates the acceptable range for $M$ and $M_B$ that accounts for the observed baryon asymmetry ($Y_B=9.2\times 10^{-11}$) for $\gamma=3$, $n=1.3$, and various values of $\alpha$. These plots correspond to a dimension-6 B-violating interaction ($m=2$).

The solid black curve represents a baryon-to-entropy ratio of $Y_B = 9.2 \times 10^{-11}$, while the red dashed curve and the blue solid line correspond to baryon-to-entropy ratios of $Y_B = 8.8 \times 10^{-11}$ and $Y_B = 9.8 \times 10^{-11}$, respectively. 

Let us now concentrate on Starobinsky's model, characterized by the function $F(R)=R+{R^2}/{6M^2}$, where $M$ represents a mass dimension. In this model, the $R^2$-term becomes significant during inflation, overshadowing the effects of relativistic matter. Consequently, the power-law scale factor scenario within the Friedmann-Robertson-Walker (FRW) metric, when applied to Starobinsky's model, fails to adequately account for inflation and, subsequently, gravitational baryogenesis.

\begin{center}
\begin{tabular}[t]{|c|c|c|c|c|c|c|c|c|}
\hline
\multicolumn{9}{|c|}{Table 1: Evaluation of the model's parameters} \\
\hline
$n$ &$m$ & $\alpha$ & $\gamma$ & $M$ & $M_B$ &$\tau_D$ & $Y_B$& \\
\hline
$1$ & $2$ & $0$ & $1$ & $10^{-8}{M_P}$&$10^{-4}M_p$& $1.67\times10^{13}Gev$ & $2.8\times10^{-14}$& \\
\hline
$1$ & $2$ & $0$ & $2$ & $10^{-8}{M_P}$&$10^{-4}M_p$& $1.67\times10^{13}Gev$ & $2.1\times10^{-14}$& \\
\hline
$1$ & $2$ & $0$ & $3$ & $10^{-8}{M_P}$&$10^{-4}M_p$& $1.67\times10^{13}Gev$ & $1.59\times10^{-14}$& \\
\hline
$1$ & $2$ & $1$ & $3$ & $10^{-8}{M_P}$&$10^{-4}M_p$& $2.79\times10^{13}Gev$ & $1.52\times10^{-12}$& \\
\hline
$1$ & $2$ & $2$ & $3$ & $10^{-8}{M_P}$&$10^{-4}M_p$& $3.15\times10^{13}Gev$ & $4.39\times10^{-12}$& \\
\hline
$1$ & $2$ & $3$ & $3$ & $10^{-8}{M_P}$&$10^{-4}M_p$& $3.32\times10^{13}Gev$ & $5.91\times10^{-12}$& \\
\hline
$1.4$ & $2$ & $3$ & $3$ & $10^{-8}{M_P}$&$10^{-4}M_p$& $3.55\times10^{12}Gev$ & $5.14\times10^{-10}$& \\
\hline
$1.4$ & $2$ & $3$ & $3$ & $10^{-8}{M_P}$&$10^{-5}M_p$& $4.\times10^{12}Gev$ & $1.56\times10^{-10}$& \\
\hline
$1.4$ & $2$ & $3$ & $3$ & $10^{-7}{M_P}$&$10^{-4}M_p$& $5.16\times10^{13}Gev$ & $1.61\times10^{-9}$& \\
\hline
\end{tabular}
\end{center}

\section{Conclusion}

In this paper, we investigated the gravitational baryogenesis mechanism in the $F(R)$ gravity's rainbow.
We used coupling between derivative of Ricci scalar curvature and baryon current to describe baryon asymmetry.
We have chosen power-law type of $F(R)$ model: $F(R)=BR^n$ with $B$ and $n$ being arbitrary real constant. Additionally, the rainbow function is expressed as a power-law form of the Hubble parameter: $\tilde{f}\approx H^\alpha/M^\alpha$ with $M$ and $\alpha$ also being arbitrary real constant. We find power-law solution in the form $a(t)=A t^\gamma$ for this model.
A key challenge is identifying suitable parameters for the model that satisfy its constraints and maintain consistency between predicted and observed values.

The main constraints are $\{\rho, R\} > 0$ and $H^\alpha \gg M^\alpha$.
 We identify the acceptable range for the model parameters through numerical analysis. 
We conclude that for $n\geq2$, there are no solutions for the expansion of the universe and gravitational baryogenesis. The plausible solutions for values close to $n=2$ involve very large values of $\alpha$ or $\gamma$. 
If $n=1.5$ and $\alpha = 0.01$, then $\gamma$ must be greater than 2.
Similarly, if $n=1.4$ and $\alpha = 3$, $\gamma$ must also be greater than 2. We consider the evolution of decoupling temperature $T_D$ and baryon-to-entropy ratio as function of the model's parameters. 
Our graphs indicate that the ratio of baryon-to-entropy in relation to $\gamma$ reaches a maximum at $\gamma_{max}$ for specific, acceptable values of $\alpha$, $n$, and $m$. Additionally, as the values of $\alpha$ increase, the maximum value of this ratio decreases.
We analyze the predictions of this model in relation to current observational data. Our findings indicate that $F(R)$ gravity’s rainbow offers a unique approach to gravitational baryogenesis, which may account for the observed baryon asymmetry in the universe while also integrating quantum gravity corrections.

\bibliography{baryoge}{}

\begin{thebibliography}{49}
\providecommand{\natexlab}[1]{#1}
\providecommand{\url}[1]{\texttt{#1}}
\expandafter\ifx\csname urlstyle\endcsname\relax
  \providecommand{\doi}[1]{doi: #1}\else
  \providecommand{\doi}{doi: \begingroup \urlstyle{rm}\Url}\fi

\bibitem[Affleck and Dine(1985)]{Affleck:1984fy}
Ian Affleck and Michael Dine.
\newblock {A New Mechanism for Baryogenesis}.
\newblock \emph{Nucl. Phys. B}, 249:\penalty0 361--380, 1985.
\newblock \doi{10.1016/0550-3213(85)90021-5}.

\bibitem[Aghanim et~al.(2020)]{Planck:2018vyg}
N.~Aghanim et~al.
\newblock {Planck 2018 results. VI. Cosmological parameters}.
\newblock \emph{Astron. Astrophys.}, 641:\penalty0 A6, 2020.
\newblock \doi{10.1051/0004-6361/201833910}.
\newblock [Erratum: Astron.Astrophys. 652, C4 (2021)].

\bibitem[Amelino-Camelia(2013)]{Amelino-Camelia:2008aez}
Giovanni Amelino-Camelia.
\newblock {Quantum-Spacetime Phenomenology}.
\newblock \emph{Living Rev. Rel.}, 16:\penalty0 5, 2013.
\newblock \doi{10.12942/lrr-2013-5}.

\bibitem[Ashour et~al.(2019)Ashour, Alcheikh, and Chamoun]{Ashour:2019wup}
A.~Ashour, M.~Alcheikh, and N.~Chamoun.
\newblock {General Modified Friedmann Equations in Rainbow Flat Universe, by
  Thermodynamics}.
\newblock \emph{Eur. Phys. J. C}, 79\penalty0 (2):\penalty0 127, 2019.
\newblock \doi{10.1140/epjc/s10052-019-6640-8}.

\bibitem[Awad et~al.(2013)Awad, Ali, and Majumder]{Awad:2013nxa}
Adel Awad, Ahmed~Farag Ali, and Barun Majumder.
\newblock {Nonsingular Rainbow Universes}.
\newblock \emph{JCAP}, 10:\penalty0 052, 2013.
\newblock \doi{10.1088/1475-7516/2013/10/052}.

\bibitem[Azhar et~al.(2020)Azhar, Jawad, and Rani]{Azhar:2020coz}
Nadeem Azhar, Abdul Jawad, and Shamaila Rani.
\newblock {Generalized gravitational baryogenesis of well-known $f(T,TG)$ and
  $f(T,B)$ models}.
\newblock \emph{Phys. Dark Univ.}, 30:\penalty0 100724, 2020.
\newblock \doi{10.1016/j.dark.2020.100724}.

\bibitem[Baffou et~al.(2019)Baffou, Houndjo, Kanfon, and
  Salako]{Baffou:2018hpe}
E.~H. Baffou, M.~J.~S. Houndjo, D.~A. Kanfon, and I.~G. Salako.
\newblock {$f(R,T)$ models applied to baryogenesis}.
\newblock \emph{Eur. Phys. J. C}, 79\penalty0 (2):\penalty0 112, 2019.
\newblock \doi{10.1140/epjc/s10052-019-6559-0}.

\bibitem[Bennett et~al.(2003)]{WMAP:2003ivt}
C.~L. Bennett et~al.
\newblock {First year Wilkinson Microwave Anisotropy Probe (WMAP) observations:
  Preliminary maps and basic results}.
\newblock \emph{Astrophys. J. Suppl.}, 148:\penalty0 1--27, 2003.
\newblock \doi{10.1086/377253}.

\bibitem[Bhattacharjee and Sahoo(2020)]{Bhattacharjee:2020wbh}
Snehasish Bhattacharjee and P.~K. Sahoo.
\newblock {Baryogenesis in $f(Q,{\mathcal { T}})$ gravity}.
\newblock \emph{Eur. Phys. J. C}, 80\penalty0 (3):\penalty0 289, 2020.
\newblock \doi{10.1140/epjc/s10052-020-7844-7}.

\bibitem[Burton-Villalobos et~al.(2025)Burton-Villalobos, Otalora,
  Gonzalez-Espinoza, and Leyva]{Burton-Villalobos:2025tgc}
Andr\'es Burton-Villalobos, Giovanni Otalora, Manuel Gonzalez-Espinoza, and
  Yoelsy Leyva.
\newblock {Classical and quantum cosmology of $f(R)$ gravity's rainbow in
  Schutz's formalism}.
\newblock 2 2025.

\bibitem[Capozziello and De~Laurentis(2011)]{Capozziello:2011et}
Salvatore Capozziello and Mariafelicia De~Laurentis.
\newblock {Extended Theories of Gravity}.
\newblock \emph{Phys. Rept.}, 509:\penalty0 167--321, 2011.
\newblock \doi{10.1016/j.physrep.2011.09.003}.

\bibitem[Channuie(2019)]{Channuie:2019kus}
Phongpichit Channuie.
\newblock {Deformed Starobinsky model in gravity\textquoteright{}s rainbow}.
\newblock \emph{Eur. Phys. J. C}, 79\penalty0 (6):\penalty0 508, 2019.
\newblock \doi{10.1140/epjc/s10052-019-7031-x}.

\bibitem[Chatrabhuti et~al.(2016)Chatrabhuti, Yingcharoenrat, and
  Channuie]{Chatrabhuti:2015mws}
Auttakit Chatrabhuti, Vicharit Yingcharoenrat, and Phongpichit Channuie.
\newblock {Starobinsky Model in Rainbow Gravity}.
\newblock \emph{Phys. Rev. D}, 93\penalty0 (4):\penalty0 043515, 2016.
\newblock \doi{10.1103/PhysRevD.93.043515}.

\bibitem[Cohen and Kaplan(1987)]{Cohen:1987vi}
Andrew~G. Cohen and David~B. Kaplan.
\newblock {Thermodynamic Generation of the Baryon Asymmetry}.
\newblock \emph{Phys. Lett. B}, 199:\penalty0 251--258, 1987.
\newblock \doi{10.1016/0370-2693(87)91369-4}.

\bibitem[Cohen et~al.(1991)Cohen, Kaplan, and Nelson]{Cohen:1991iu}
Andrew~G. Cohen, D.~B. Kaplan, and A.~E. Nelson.
\newblock {Spontaneous baryogenesis at the weak phase transition}.
\newblock \emph{Phys. Lett. B}, 263:\penalty0 86--92, 1991.
\newblock \doi{10.1016/0370-2693(91)91711-4}.

\bibitem[Cohen et~al.(1998)Cohen, De~Rujula, and Glashow]{Cohen:1997ac}
Andrew~G. Cohen, A.~De~Rujula, and S.~L. Glashow.
\newblock {A Matter - antimatter universe?}
\newblock \emph{Astrophys. J.}, 495:\penalty0 539--549, 1998.
\newblock \doi{10.1086/305328}.

\bibitem[Cyburt et~al.(2016)Cyburt, Fields, Olive, and Yeh]{Cyburt:2015mya}
Richard~H. Cyburt, Brian~D. Fields, Keith~A. Olive, and Tsung-Han Yeh.
\newblock {Big Bang Nucleosynthesis: 2015}.
\newblock \emph{Rev. Mod. Phys.}, 88:\penalty0 015004, 2016.
\newblock \doi{10.1103/RevModPhys.88.015004}.

\bibitem[Davoudiasl et~al.(2004)Davoudiasl, Kitano, Kribs, Murayama, and
  Steinhardt]{Davoudiasl:2004gf}
Hooman Davoudiasl, Ryuichiro Kitano, Graham~D. Kribs, Hitoshi Murayama, and
  Paul~J. Steinhardt.
\newblock {Gravitational baryogenesis}.
\newblock \emph{Phys. Rev. Lett.}, 93:\penalty0 201301, 2004.
\newblock \doi{10.1103/PhysRevLett.93.201301}.

\bibitem[De~Felice and Tsujikawa(2010)]{DeFelice:2010aj}
Antonio De~Felice and Shinji Tsujikawa.
\newblock {f(R) theories}.
\newblock \emph{Living Rev. Rel.}, 13:\penalty0 3, 2010.
\newblock \doi{10.12942/lrr-2010-3}.

\bibitem[Fukugita and Yanagida(1986)]{Fukugita:1986hr}
M.~Fukugita and T.~Yanagida.
\newblock {Baryogenesis Without Grand Unification}.
\newblock \emph{Phys. Lett. B}, 174:\penalty0 45--47, 1986.
\newblock \doi{10.1016/0370-2693(86)91126-3}.

\bibitem[Fukushima et~al.(2016)Fukushima, Mizuno, and Maeda]{Fukushima:2016wyz}
Mitsuhiro Fukushima, Shuntaro Mizuno, and Kei-ichi Maeda.
\newblock {Gravitational Baryogenesis after Anisotropic Inflation}.
\newblock \emph{Phys. Rev. D}, 93\penalty0 (10):\penalty0 103513, 2016.
\newblock \doi{10.1103/PhysRevD.93.103513}.

\bibitem[Gim et~al.(2018)Gim, Um, and Kim]{Gim:2017rmn}
Yongwan Gim, Hwajin Um, and Wontae Kim.
\newblock {Black hole complementarity with the generalized uncertainty
  principle in gravity's rainbow}.
\newblock \emph{JCAP}, 02:\penalty0 060, 2018.
\newblock \doi{10.1088/1475-7516/2018/02/060}.

\bibitem[Goodarzi(2023)]{Goodarzi:2023ltp}
Parviz Goodarzi.
\newblock {Gravitational baryogenesis in non-minimal kinetic coupling model}.
\newblock \emph{Eur. Phys. J. C}, 83\penalty0 (11):\penalty0 990, 2023.
\newblock \doi{10.1140/epjc/s10052-023-12182-7}.

\bibitem[Gorji et~al.(2017)Gorji, Nozari, and Vakili]{Gorji:2016laj}
M.~A. Gorji, K.~Nozari, and B.~Vakili.
\newblock {Gravity's rainbow: a bridge between LQC and DSR}.
\newblock \emph{Phys. Lett. B}, 765:\penalty0 113--119, 2017.
\newblock \doi{10.1016/j.physletb.2016.12.023}.

\bibitem[Kolb and Turner(2019)]{Kolb:1990vq}
Edward~W. Kolb and Michael~S. Turner.
\newblock \emph{{The Early Universe}}, volume~69.
\newblock Taylor and Francis, 5 2019.
\newblock ISBN 978-0-429-49286-0, 978-0-201-62674-2.
\newblock \doi{10.1201/9780429492860}.

\bibitem[Lambiase and Scarpetta(2006)]{Lambiase:2006dq}
Gaetano Lambiase and G.~Scarpetta.
\newblock {Baryogenesis in f(R): Theories of Gravity}.
\newblock \emph{Phys. Rev. D}, 74:\penalty0 087504, 2006.
\newblock \doi{10.1103/PhysRevD.74.087504}.

\bibitem[Leyva and Otalora(2023)]{Leyva:2022zhz}
Yoelsy Leyva and Giovanni Otalora.
\newblock {Revisiting f(R) gravity's rainbow: Inflation and primordial
  fluctuations}.
\newblock \emph{JCAP}, 04:\penalty0 030, 2023.
\newblock \doi{10.1088/1475-7516/2023/04/030}.

\bibitem[Lima and Singleton(2016)]{Lima:2016cbh}
J.~A.~S. Lima and Douglas Singleton.
\newblock {The Impact of Particle Production on Gravitational Baryogenesis}.
\newblock \emph{Phys. Lett. B}, 762:\penalty0 506--511, 2016.
\newblock \doi{10.1016/j.physletb.2016.10.005}.

\bibitem[Ling(2007)]{Ling:2006az}
Yi~Ling.
\newblock {Rainbow universe}.
\newblock \emph{JCAP}, 08:\penalty0 017, 2007.
\newblock \doi{10.1088/1475-7516/2007/08/017}.

\bibitem[Magueijo and Smolin(2004)]{Magueijo:2002xx}
Joao Magueijo and Lee Smolin.
\newblock {Gravity's rainbow}.
\newblock \emph{Class. Quant. Grav.}, 21:\penalty0 1725--1736, 2004.
\newblock \doi{10.1088/0264-9381/21/7/001}.

\bibitem[Nojiri and Odintsov(2011)]{Nojiri:2010wj}
Shin'ichi Nojiri and Sergei~D. Odintsov.
\newblock {Unified cosmic history in modified gravity: from F(R) theory to
  Lorentz non-invariant models}.
\newblock \emph{Phys. Rept.}, 505:\penalty0 59--144, 2011.
\newblock \doi{10.1016/j.physrep.2011.04.001}.

\bibitem[Nozari and Rajabi(2018)]{Nozari:2018ift}
Kourosh Nozari and Fateme Rajabi.
\newblock {Baryogenesis in $f(R,T)$ Gravity}.
\newblock \emph{Commun. Theor. Phys.}, 70\penalty0 (4):\penalty0 451, 2018.
\newblock \doi{10.1088/0253-6102/70/4/451}.

\bibitem[Odintsov and Oikonomou(2016)]{Odintsov:2016hgc}
S.~D. Odintsov and V.~K. Oikonomou.
\newblock {Gauss\textendash{}Bonnet gravitational baryogenesis}.
\newblock \emph{Phys. Lett. B}, 760:\penalty0 259--262, 2016.
\newblock \doi{10.1016/j.physletb.2016.06.074}.

\bibitem[Oikonomou and Saridakis(2016)]{Oikonomou:2016jjh}
V.~K. Oikonomou and Emmanuel~N. Saridakis.
\newblock {$f(T)$ gravitational baryogenesis}.
\newblock \emph{Phys. Rev. D}, 94\penalty0 (12):\penalty0 124005, 2016.
\newblock \doi{10.1103/PhysRevD.94.124005}.

\bibitem[Ramos and P\'aramos(2017)]{Ramos:2017cot}
M.~P. L.~P. Ramos and J.~P\'aramos.
\newblock {Baryogenesis in Nonminimally Coupled $f(R)$ Theories}.
\newblock \emph{Phys. Rev. D}, 96\penalty0 (10):\penalty0 104024, 2017.
\newblock \doi{10.1103/PhysRevD.96.104024}.

\bibitem[Rovelli(2008)]{Rovelli:2008cj}
Carlo Rovelli.
\newblock {A Note on DSR}.
\newblock 8 2008.

\bibitem[Sadjadi(2007)]{Sadjadi:2007dx}
H.~Mohseni Sadjadi.
\newblock {A Note on Gravitational Baryogenesis}.
\newblock \emph{Phys. Rev. D}, 76:\penalty0 123507, 2007.
\newblock \doi{10.1103/PhysRevD.76.123507}.

\bibitem[Sahoo and Bhattacharjee(2020)]{Sahoo:2019pat}
P.~K. Sahoo and Snehasish Bhattacharjee.
\newblock {Gravitational Baryogenesis in Non-Minimal Coupled $f(R,T)$ Gravity}.
\newblock \emph{Int. J. Theor. Phys.}, 59\penalty0 (5):\penalty0 1451--1459,
  2020.
\newblock \doi{10.1007/s10773-020-04414-3}.

\bibitem[Sakharov(1967)]{Sakharov:1967dj}
A.~D. Sakharov.
\newblock {Violation of CP Invariance, C asymmetry, and baryon asymmetry of the
  universe}.
\newblock \emph{Pisma Zh. Eksp. Teor. Fiz.}, 5:\penalty0 32--35, 1967.
\newblock \doi{10.1070/PU1991v034n05ABEH002497}.

\bibitem[Sefiedgar and Daghigh(2017)]{Sefiedgar:2017tnt}
Akram~Sadat Sefiedgar and Majid Daghigh.
\newblock {Thermodynamics of the FRW universe in rainbow gravity}.
\newblock \emph{Int. J. Mod. Phys. D}, 26\penalty0 (13):\penalty0 1750139,
  2017.
\newblock \doi{10.1142/S0218271817501395}.

\bibitem[Shahjalal(2018)]{Shahjalal:2018hid}
Md. Shahjalal.
\newblock {Phase transition of quantum-corrected Schwarzschild black hole in
  rainbow gravity}.
\newblock \emph{Phys. Lett. B}, 784:\penalty0 6--11, 2018.
\newblock \doi{10.1016/j.physletb.2018.07.032}.

\bibitem[Sotiriou and Faraoni(2010)]{Sotiriou:2008rp}
Thomas~P. Sotiriou and Valerio Faraoni.
\newblock {f(R) Theories Of Gravity}.
\newblock \emph{Rev. Mod. Phys.}, 82:\penalty0 451--497, 2010.
\newblock \doi{10.1103/RevModPhys.82.451}.

\bibitem[Spergel et~al.(2007)]{WMAP:2006bqn}
D.~N. Spergel et~al.
\newblock {Wilkinson Microwave Anisotropy Probe (WMAP) three year results:
  implications for cosmology}.
\newblock \emph{Astrophys. J. Suppl.}, 170:\penalty0 377, 2007.
\newblock \doi{10.1086/513700}.

\bibitem[Stelle(1977)]{Stelle:1976gc}
K.~S. Stelle.
\newblock {Renormalization of Higher Derivative Quantum Gravity}.
\newblock \emph{Phys. Rev. D}, 16:\penalty0 953--969, 1977.
\newblock \doi{10.1103/PhysRevD.16.953}.

\bibitem[Upadhyay et~al.(2018)Upadhyay, Hendi, Panahiyan, and
  Eslam~Panah]{Upadhyay:2018vfu}
Sudhaker Upadhyay, Seyed~Hossein Hendi, Shahram Panahiyan, and Behzad
  Eslam~Panah.
\newblock {Thermal fluctuations of charged black holes in
  gravity\textquoteright{}s rainbow}.
\newblock \emph{PTEP}, 2018\penalty0 (9):\penalty0 093E01, 2018.
\newblock \doi{10.1093/ptep/pty093}.

\bibitem[Usman et~al.(2024)Usman, Jawad, and Sultan]{Usman:2024cya}
Muhammad Usman, Abdul Jawad, and Abdul~Malik Sultan.
\newblock {Compatibility of gravitational baryogenesis in $f(Q, C)$ gravity}.
\newblock \emph{Eur. Phys. J. C}, 84\penalty0 (8):\penalty0 868, 2024.
\newblock \doi{10.1140/epjc/s10052-024-13219-1}.

\bibitem[Waeming and Channuie(2020)]{Waeming:2020rir}
Areef Waeming and Phongpichit Channuie.
\newblock {Inflation from f(R) theories in gravity\textquoteright{}s rainbow}.
\newblock \emph{Eur. Phys. J. C}, 80\penalty0 (9):\penalty0 802, 2020.
\newblock \doi{10.1140/epjc/s10052-020-8387-7}.

\bibitem[Weinberg(1979)]{Weinberg:1979bt}
Steven Weinberg.
\newblock {Cosmological Production of Baryons}.
\newblock \emph{Phys. Rev. Lett.}, 42:\penalty0 850--853, 1979.
\newblock \doi{10.1103/PhysRevLett.42.850}.

\bibitem[Yuennan et~al.(2024)Yuennan, Channuie, and Momeni]{Yuennan:2024lkz}
Jureeporn Yuennan, Phongpichit Channuie, and Davood Momeni.
\newblock {Gravity\textquoteright{}s rainbow effects on higher curvature
  modification of $R^{2}$ inflation}.
\newblock \emph{Eur. Phys. J. C}, 84\penalty0 (8):\penalty0 766, 2024.
\newblock \doi{10.1140/epjc/s10052-024-13155-0}.

\end{thebibliography}
\bibliographystyle{plainnat.bst}
\end{document}